\documentclass[preprint,1p,10pt,review]{elsarticle}

\usepackage[utf8]{inputenc}
\usepackage{amsfonts}
\usepackage{bibnames}
\usepackage{float}
\usepackage{mathtools}
\usepackage{amsthm}
\usepackage{amsmath}
\numberwithin{equation}{section}
\usepackage{xfrac}
\usepackage{graphicx}
\usepackage{relsize} 
\usepackage{breqn}
\usepackage{siunitx}

\usepackage{algorithm}
\usepackage{algpseudocode}
\usepackage{wrapfig}

\usepackage{tabularx}
\usepackage{booktabs}
\usepackage{multicol}
\usepackage{multirow}
\usepackage{dcolumn}

\usepackage[margin=1cm]{caption}  
\usepackage{subcaption}
\usepackage[colorlinks,urlcolor=blue,unicode]{hyperref} 
\usepackage[capitalise]{cleveref}

\usepackage{enumitem} 
\setenumerate{topsep=6pt,itemsep=2pt,partopsep=2pt, parsep=2pt} 
\setitemize{topsep=6pt,itemsep=2pt,partopsep=2pt, parsep=2pt} 

\algnewcommand\algorithmicforeach{\textbf{for each}}
\algdef{S}[FOR]{ForEach}[1]{\algorithmicforeach\ #1\ \algorithmicdo}

\usepackage{todonotes}

\newcommand{\m}[1]{\ensuremath{\boldsymbol{#1}}}

\theoremstyle{definition}

\usepackage{empheq}
\DeclareCaptionType{equ}[][]

\usepackage{tikz}
\usepackage{pgfplots}
\usetikzlibrary{positioning}
\definecolor{lightest}{RGB}{200, 200, 200}
\definecolor{lighter}{RGB}{150, 150, 150}
\definecolor{mid}{RGB}{100, 100, 100}
\definecolor{darker}{RGB}{50, 50, 50}
\definecolor{darkest}{RGB}{0, 0, 0}

\usepackage{setspace}

\newcommand{\cicmunda}{server}

\newcommand{\cicmundove}{servers}
\newcommand{\cicmundoveb}{Servers}

\newcommand{\cp}{{\fontfamily{bch}\selectfont{\small{CP\smaller{$_\text{{0}}$}}}}}

\newcommand{\cpwse}{{\fontfamily{bch}\selectfont{\small{CP\smaller{$_{\text{WS-SE}}$}}}}}
\newcommand{\cpwseic}{{\fontfamily{bch}\selectfont{\small{CP\smaller{$_{\text{WS-SEIC}}$}}}}}

\newcommand{\lss}{{\fontfamily{bch}\selectfont{\small{LOSOS}}}}
\newcommand{\lsse}{{\fontfamily{bch}\selectfont{\small{LOSOS\smaller{$_{\text{SE}}$}}}}}
\newcommand{\lsseic}{{\fontfamily{bch}\selectfont{\small{LOSOS\smaller{$_{\text{SEIC}}$}}}}}
\newcommand{\rsl}{{\fontfamily{bch}\selectfont{\small{ROSOL}}}}
\newcommand{\rsle}{{\fontfamily{bch}\selectfont{\small{ROSOL\smaller{$_{\text{SE}}$}}}}}
\newcommand{\rsleic}{{\fontfamily{bch}\selectfont{\small{ROSOL\smaller{$_{\text{SEIC}}$}}}}}

\newcommand{\taskSetEq}{T}
\newcommand{\taskOneIndexEq}{i}
\newcommand{\taskTwoIndexEq}{j}
\newcommand{\taskOneEq}{\taskSetEq_{\taskOneIndexEq}}
\newcommand{\taskTwoEq}{\taskSetEq_{\taskTwoIndexEq}}

\newcommand{\taskSet}{$\taskSetEq$}

\newcommand{\taskOne}{$\taskOneEq$}
\newcommand{\taskTwo}{$\taskTwoEq$}
\newcommand{\taskLast}{T_{t}}

\newcommand{\taskOneProcessEq}{p_{\taskOneIndexEq}}
\newcommand{\taskOneStart}{$s_{\taskOneIndexEq}$}
\newcommand{\taskOneProcess}{$\taskOneProcessEq$}
\newcommand{\taskOneCompletion}{$c_{\taskOneIndexEq}$}
\newcommand{\taskTwoStart}{$s_{\taskTwoIndexEq}$}

\newcommand{\machineSetEq}{M}
\newcommand{\machineSet}{$\machineSetEq$}
\newcommand{\machineEq}{M_{k}}
\newcommand{\machine}{$\machineEq$}
\newcommand{\machineLast}{M_{m}}

\newcommand{\machineLong}{$M_{longest}$}

\newcommand{\machineLastIndex}{$m$}

\newcommand{\workerSet}{R}
\newcommand{\workerLast}{R_{r}}
\newcommand{\worker}{$R_{x}$}

\newcommand{\workerLastIndex}{r}

\newcommand{\setupMatrixEq}{\m{O}}
\newcommand{\setupMatrix}{$\setupMatrixEq$}
\newcommand{\setupTimeEq}{o_{i,j}}
\newcommand{\setupTime}{$\setupTimeEq$}
\newcommand{\setupMatrixDefined}{$\setupMatrixEq = [\setupTimeEq] \in (\mathbb{N} \cup \{\infty\}) ^{{t}\times{t}}$}
\newcommand{\usedSetupSet}{$U$}
\newcommand{\usedSetupOne}{$U_{y}$}
\newcommand{\usedSetupTwo}{$U_{z}$}
\newcommand{\usedSetupLast}{$U_{u}$}

\newcommand{\usedSetupOneStart}{$\overline{s}_{y}$}
\newcommand{\usedSetupOneProcess}{$\overline{p}_{y}$}
\newcommand{\usedSetupOneCompletion}{$\overline{c}_{y}$}
\newcommand{\usedSetupTwoStart}{$\overline{s}_{z}$}

\newcommand{\usedSetupTwoCompletion}{$\overline{c}_{z}$}
\newcommand{\usedSetupFromSet}{$U_{y} \in U$}

\newcommand{\taskIntervalSetEq}{I^{T}}
\newcommand{\taskOneIntervalEq}{I^{T}_{i}}

\newcommand{\taskOneMachineIntervalEq}{I^{T^{Opt}}_{i,k}}

\newcommand{\taskIntervalSet}{$\taskIntervalSetEq$}
\newcommand{\taskOneInterval}{$\taskOneIntervalEq$}

\newcommand{\taskOneMachineInterval}{$\taskOneMachineIntervalEq$}

\newcommand{\generalSetupIntervalSetEq}{I^{S}}
\newcommand{\generalSetupIntervalSet}{$\generalSetupIntervalSetEq$}
\newcommand{\generalSetupIntervalEq}{I^{S}_{i}}
\newcommand{\generalSetupInterval}{$\generalSetupIntervalEq$}

\newcommand{\optimizingCriterion}{$C_{max}$}

\newcommand{\inN}{$\in \mathbb{N}$}
\newcommand{\inNZero}{$\in \mathbb{N}_{0}$}
\newcommand{\NPHard}{$\mathcal{NP}$-hard}

\newcommand{\cpLengthOf}{\textsc{LengthOf}}
\newcommand{\cpStartOf}{\textsc{StartOf}}
\newcommand{\cpEndOf}{\textsc{EndOf}}
\newcommand{\cpNoOverlap}{\textsc{NoOverlap}}
\newcommand{\cpAlternative}{\textsc{Alternative}}

\newcommand{\cpPulse}{\textsc{Pulse}}
\newcommand{\cpStartOfNext}{\textsc{StartOfNext}}
\newcommand{\cpEndAtStart}{\textsc{EndAtStart}}

\usepackage{amssymb}

\usepackage{lineno}

\journal{Computers \& Industrial Engineering}

\pgfplotsset{compat=1.17} 

\begin{document}
\usetikzlibrary{arrows.meta, calc, chains, quotes, positioning, shapes.geometric}

\begin{frontmatter}



\title{Constraint Programming and Constructive Heuristics for Parallel Machine Scheduling with Sequence-Dependent Setups and Common Servers}


\author[fee,ciirc]{Vilém Heinz\corref{cor1}}
\ead{vilem.heinz@cvut.cz}
\author[fee,ciirc]{Antonín Novák}
\author[ciirc]{Marek Vlk}
\author[ciirc]{Zdeněk Hanzálek}
\address[fee]{Faculty of Electrical Engineering, \\Czech Technical University in Prague, CZ}
\address[ciirc]{Czech Institute of Informatics, Robotics and Cybernetics, \\Czech Technical University in Prague, CZ}
\cortext[cor1]{Corresponding author}



\begin{abstract}
This paper examines scheduling problem denoted as $P|seq, ser|C_{max}$ in Graham's notation; in other words, scheduling of tasks on parallel identical machines ($P$) with sequence-dependent setups ($seq$) each performed by one of the available \cicmundove\ ($ser$). The goal is to minimize the makespan ($C_{max}$). We propose a Constraint Programming (CP) model for finding the optimal solution and constructive heuristics suitable for large problem instances. These heuristics are also used to provide a feasible starting solution to the proposed CP model, significantly improving its efficiency. This combined approach constructs solutions for benchmark instances of up to 20 machines and 500 tasks in 10 seconds, with makespans \SIrange{3}{11.5}{\percent}
greater than the calculated lower bounds with a 5\% average. The extensive experimental comparison also shows that our proposed approaches outperform the existing ones.
\end{abstract}


\begin{highlights}
\item Effective Constraint Programming model is proposed.
\item Constructive heuristics with multiple improvement methods are presented.
\item An effective approach composed of Constraint Programming and heuristics is proposed.
\item Improvement over the state-of-the-art exact and heuristic approaches is achieved.
\end{highlights}


\end{frontmatter}


\section{Introduction}
\label{sec:intro}
In recent years, manufacturing has become more complex than ever. Factories must handle a large variety of products on multiple production lines, customers demand highly proprietary products of small-batch volumes, and companies are constantly forced to adapt to market shifts. We are in an environment where the organization of work and effective utilization of production capacities is a crucial yet very complex problem to tackle \cite{_2017ijat}.

In this paper, we focus on a scenario often encountered in real-world production. Imagine a production environment with multiple identical machines. Tasks are to be freely assigned to these machines. Depending on their assignment and order, adjustments on said machines (sequence-dependent setups) are performed between them by \cicmundove. The number of \cicmundove\ can be arbitrary, and any \cicmunda\ can perform any setup.

\paragraph{Example} To illustrate the importance of the considered problem, we provide an example of a production planning challenge tackled by an existing company. This company produces plastic tubes of different shapes and sizes but from the same material and with a very similar manufacturing process. Thus, all machines they use are the same and depending only on their settings, they produce plastic tubes of different shapes and/or sizes. These machines' settings are adjusted by workers present in the factory. Since every tube shape and size requires different machine settings to be produced, adjustment between each tube pair in production has its own unique sequence and can require variable time to adjust. Clearly, there is no need to adjust the machine between the production of tubes of the same shape and size.

In general, this scheduling problem containing \cicmundove, parallel identical machines and sequence-dependent setups is encountered at paint shops, printing houses, custom component manufacturing and 3D printing, foods, chemical, and other production industries where the machines are used for multiple types of different products with the necessity of adjustments. We refer the reader to \emph{Huang et al.}~\cite{hcz}, \emph{Kim et al.}~\cite{kl2} and \emph{Hamzadayi et al.}~\cite{hy} for similar problem examples.

Needless to say, human resource capacity in such productions is sometimes disregarded during the scheduling phase because of its complex real-world restrictions and uncertainties, leaving decisions to expert knowledge. We argue that even in such cases, a solution with human resource capacity taken into account can help to guide this decision process and improve the overall solution.

\subsection{Approach}
\label{sec:intro:approach}
We propose an efficient Constraint Programming (CP) \cite{ROSSI20063} model as an exact approach to the considered problem. As demonstrated in \emph{Laborie et al.}~\cite{Laborie2018}, CP is more effective than conventionally used Integer Linear Programming (ILP) for many scheduling problems. However, since the problem is \NPHard, the CP model's execution is still exponential in the worst case. Thus, we propose constructive heuristic algorithms that provide a solution in polynomial time. We also discuss several applicable improvements, yielding better solutions in exchange for more prolonged but still polynomial executions. Ultimately, we combine both approaches by using a heuristic solution as a starting point of the CP model, leveraging their distinct strengths. This combination results in significant performance improvement of the CP model, making much larger instances solvable. In general, this method of providing a starting solution to the main algorithm is called warm starting, and it was successfully applied to various scheduling problems, such as in \emph{de Abreu et al.}~\cite{DEABREU2022107976} or \emph{Pour et al.}~\cite{MPOUR2018341}.

\subsection{Contribution and Paper Outline}
\label{sec:intro:contribution}
We summarize the key contributions of the paper as:
\begin{enumerate}
    \item Addressing a new scheduling problem with \cicmundove, arising as a natural generalization of particular cases \emph{Huang et al.}~\cite{hcz}, \emph{Kim et al.}~\cite{kl2} and \emph{Hamzadayi et al.}~\cite{hy} which lay important groundwork studying aspects of our considered problem. This results in a problem combining sequence-dependent setup times and machine-task independence with extension to multiple \cicmundove\ availability.
    \item Proposing CP model utilizing cumulative resource function that can solve instances of up to ten machines and tens of tasks to the optimality usually under an hour. With a warm start, the CP model can be very effectively used on instances of tens of machines and hundreds of tasks in just several-minute runtimes.
    \item Proposing domain-specific heuristics that provide feasible solutions to instances of hundreds of machines and tens of thousands of tasks between several seconds and several minutes of runtime. With the optional improvements applied, solutions have a lower bound gap of \SIrange{3.5}{20}{\percent} with an 8\% average, measured on the benchmark instance set.
    \item Providing more efficient approaches than the state-of-the-art for related problems proposed in papers \emph{Huang et al.}~\cite{hcz}, \emph{Kim et al.}~\cite{kl2} and \emph{Hamzadayi et al.}~\cite{hy}. Both CP model as an alternative to ILP models and warm started CP model, as an alternative to state-of-the-art heuristics, show improvements in efficiency over their respective counterparts as they are able to solve instances of said related problems. This finding is very important as it proves that our approaches are not just addressing a new problem but also doing it efficiently compared to related state-of-the-art approaches.
    For our considered problem benchmark instances of sizes up to 20 machines and 500 tasks, in 10 seconds of computation, the warm started CP model provides solutions with makespans \SIrange{3}{11.5}{\percent} greater than the calculated lower bounds with a 5\% average.
\end{enumerate}

The rest of the paper is organized as follows:
\cref{sec:problem_statement} formally describes the problem and its complexity.
\cref{sec:rel_work} lists relevant literature on similar problems and discusses the current state-of-the-art approaches.
\cref{sec:cp_exact} describes the proposed CP model.
\cref{sec:constructive_heur} describes developed constructive heuristics, proposes ways of improving their solutions in a trade-off with computational time and describes constructive heuristics warm start application to the CP model.
\cref{sec:experimental_evaluation} contains the experimental results, discusses differences between proposed approaches and provides a comparison to the current state-of-the-art.
\cref{sec:conclusion} sums up the findings discovered in this paper.

\section{Problem Statement}
\label{sec:problem_statement}
The considered problem is denoted $P|seq, ser|C_{max}$ in Graham's notation. $P$ denotes parallel identical machines task scheduling, $seq$ denotes sequence-dependent setups, $ser$ denotes presence of \cicmundove\ performing said setups and $C_{max}$ denotes the overall goal of minimizing the makespan. We define problem input parameters in the following way:
\begin{itemize}
    \item Let $\machineSetEq=\{\machineSetEq_1, \ldots, \machineLast\}$ be a set of identical machines where every machine can execute at most one task or one setup at a time. The number of machines \machineLastIndex\ is given by the problem instance.
    \item Let $\workerSet=\{\workerSet_1, \ldots, \workerLast\}$ be a set of available \cicmundove. \cicmundoveb\ are considered to be identical. The number of \cicmundove\ $\workerLastIndex$ is given by the problem instance and we generally expect it to be strictly lower than \machineLastIndex.
    \item Let $\taskSetEq=\{\taskSetEq_1, \ldots, \taskLast\}$ be a set of independent non preemptive tasks where each task can be executed on any machine. Every task is described by its processing time given by the problem instance, denoted \taskOneProcess\ \inN.
    \item Let \setupMatrixDefined\ be the setup times matrix given by the problem instance, where \setupTime\ denotes the setup length between tasks \taskOne\ and \taskTwo.
\end{itemize}

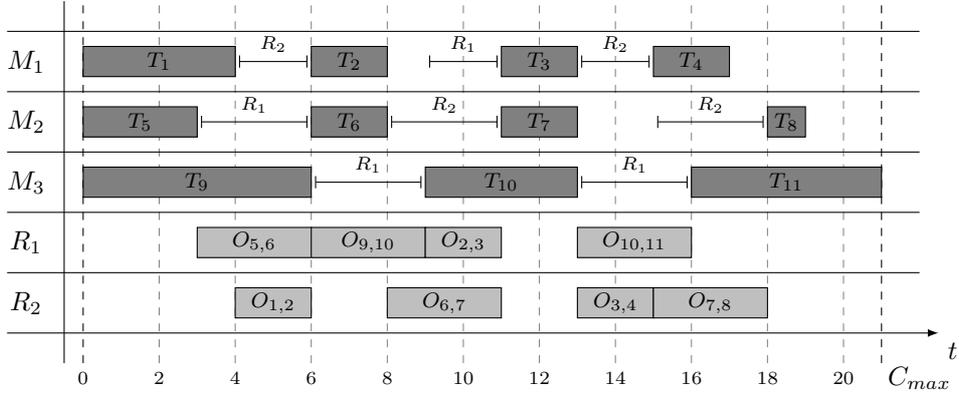
\begin{figure}[ht]
    \centering
    \captionsetup{justification=centering}
    \begin{tikzpicture}

\begin{scope}[xscale=0.5, yscale=0.4]

	\draw[-latex] (-1.5,1) -- node[anchor=north west, pos=1] {$t$} (23,1);
	\draw[] (0,0) -- (0,12);

    \begin{scope}[xshift=0.5cm]
    \draw[dashed] (0,0) -- (0,12);
    \node[] at (0,-0.5) {\scriptsize $0$};
    \draw[dashed] (21,0) -- (21,12);
    \node[] at (22,-0.5) {$C_{max}$};

    \foreach \x in {2,4,...,20}{
        \draw[dashed, help lines] (\x,0) -- (\x,12);
        \node[] at (\x,-0.5) {\scriptsize $\x$};
    }
    \end{scope}
    
    \foreach \xs/\ys/\xe/\ye in 
    	{
    	    -1.5/3/22.5/3,
    	    -1.5/5/22.5/5,
    	    -1.5/7/22.5/7,
    	    -1.5/9/22.5/9,
    	    -1.5/11/22.5/11
    	} {
        \draw (\xs,\ys) -- (\xe,\ye);
	}
	
    \node[] at (-1.0,2.0) {$R_2$};
    \node[] at (-1.0,4.0) {$R_1$};
    \node[] at (-1.0,6.0) {$M_3$};
    \node[] at (-1.0,8.0) {$M_2$};
    \node[] at (-1.0,10.0) {$M_1$};

    \foreach \xs/\ys/\xe/\ye/\lbl in 
    	{
    	    0.5/9.5/4.5/10.5/T_1, 
    	    6.5/9.5/8.5/10.5/T_2, 
    	    11.5/9.5/13.5/10.5/T_3, 
    	    15.5/9.5/17.5/10.5/T_4, 
    	    0.5/7.5/3.5/8.5/T_5, 
    	    6.5/7.5/8.5/8.5/T_6, 
    	    11.5/7.5/13.5/8.5/T_7, 
    	    18.5/7.5/19.5/8.5/T_8, 
    	    0.5/5.5/6.5/6.5/T_9, 
    	    9.5/5.5/13.5/6.5/T_{10}, 
    	    16.5/5.5/21.5/6.5/T_{11} 
    	} {
        \draw[fill=gray] (\xs,\ys) rectangle node[font=\footnotesize] {$\lbl$} (\xe,\ye);
	}
	
    \foreach \xs/\ys/\xe/\ye/\lbl in 
        {
            4.5/10/6.5/10/R_2,
            9.5/10/11.5/10/R_1,
            13.5/10/15.5/10/R_2,
            3.5/8/6.5/8/R_1,
            8.5/8/11.5/8/R_2,
            15.5/8/18.5/8/R_2,
            6.5/6/9.5/6/R_1,
            13.5/6/16.5/6/R_1
        }
    	{
            \draw[|-|, shorten <= 0.05cm, shorten >= 0.05cm] (\xs,\ys) --  node[anchor=south, font=\scriptsize, fill=white, inner sep=0, yshift=2pt, minimum width=0.3cm, minimum height=0.3cm] {$\lbl$} (\xe,\ye);
    }
	
	\foreach \xs/\ys/\xe/\ye/\lbl in 
    	{
    	    3.5/3.5/6.5/4.5/O_{5,6}, 
    	    6.5/3.5/9.5/4.5/O_{9,10}, 
    	    9.5/3.5/11.5/4.5/O_{2,3}, 
    	    13.5/3.5/16.5/4.5/O_{10,11}, 
    	    4.5/1.5/6.5/2.5/O_{1,2}, 
    	    8.5/1.5/11.5/2.5/O_{6,7}, 
    	    13.5/1.5/15.5/2.5/O_{3,4}, 
    	    15.5/1.5/18.5/2.5/O_{7,8} 
    	} {
        \draw[fill=lightgray] (\xs,\ys) rectangle node[font=\footnotesize] {$\lbl$} (\xe,\ye);
	}
	
\end{scope}
	
\end{tikzpicture}
    \caption{A Gantt chart of identical parallel machine production with sequence-dependent setups performed by \cicmundove.}
    \label{illustrative_example}
\end{figure}

The problem solution is defined by task assignments to machines and the order of tasks on each machine. For every task \taskOne\ in the problem solution, let \taskOneStart\ \inNZero\ be start time and \taskOneCompletion\ \inN\ completion time of that task in the resulting schedule. Then, let us define set of present setups in the schedule as \usedSetupSet$=\{$\usedSetupSet$_1, \ldots, $\usedSetupLast$\}$. Let \usedSetupOneStart\ \inN\ be the start, \usedSetupOneProcess\ \inN\ the processing time and \usedSetupOneCompletion\ \inN\ the completion time of the setup \usedSetupOne\ in the resulting schedule.

We consider the solution feasible if the following conditions hold:
\begin{enumerate}[label=(A\arabic*)]
    \item Every task must be executed without preemption, i.e., it cannot be temporarily suspended during the execution, so \taskOneStart\ + \taskOneProcess\ = \taskOneCompletion\ must hold.
    \item If task \taskTwo\ is next after \taskOne\ in the machine sequence, setup \setupTime\ must be performed by a \cicmunda\ and \taskTwoStart\ - \taskOneCompletion\ $\ge$ \setupTime\ must hold. This also means that no setup must be executed before the first task on each machine.
    \item Setup must be performed by exactly one \cicmunda.
    \item Setups are non-preemptive, \usedSetupOneStart\ + \usedSetupOneProcess\ = \usedSetupOneCompletion\ must hold for any \usedSetupFromSet.
    \item If \usedSetupOne\ is the setup between tasks \taskOne\ and \taskTwo\ then \taskOneCompletion\ $\le$ \usedSetupOneStart\ and \usedSetupOneCompletion\ $\le$ \taskTwoStart\ must hold.
    \item For any two setups \usedSetupOne, \usedSetupTwo\ performed by the same \worker, \usedSetupTwoStart\ $\ge$ \usedSetupOneCompletion\ or \usedSetupOneStart\ $\ge$ \usedSetupTwoCompletion\ must hold. This also ensures that \cicmunda\ can perform at most one setup at a time.
\end{enumerate}

Finally, the makespan minimization can be thought of as the minimization of the latest task completion time.

\cref{illustrative_example} shows an example of feasible schedule of the considered problem. In this case, we have 3 machines (\machineSet$_{1}$ to \machineSet$_{3}$), 2 \cicmundove\ ($\workerSet_{1}$ and $\workerSet_{2}$) and 11 tasks (\taskSet$_{1}$ to \taskSet$_{11}$). Each row in \cref{illustrative_example} represents the schedule of one problem's resource, either a machine or a \cicmunda. Setups on machines are marked by \cicmunda\ performing them. The zero represents the start and \optimizingCriterion\ the end of the schedule, i.e. makespan. Task processing times and their respective setups in the schedule are in \cref{example_times}. The makespan is 21.

\begin{table}[h]
    {
        \centering
        \begin{tabular}{|c|c|c|c|c|c|c|c|c|c|c|}
            \hline
                $T_1$ & $T_2$ & $T_3$ & $T_4$ & $T_5$ & $T_6$ & $T_7$ & $T_8$ & $T_9$ & $T_{10}$ & $T_{11}$\\
            \hline
                4 & 2 & 2 & 2 & 3 & 2 & 2 & 1 & 6 & 4 & 5\\
            \hline
        \end{tabular}
        
        \vspace{4pt}
        
        \centering
        \begin{tabular}{|c|c|c|c|c|c|c|c|}
            \hline
                $O_{5,6}$ & $O_{9,10}$ & $O_{2,3}$ & $O_{10,11}$ & $O_{1,2}$ & $O_{6,7}$ & $O_{3,4}$ & $O_{7,8}$\\
            \hline
                3 & 3 & 2 & 3 & 2 & 3 & 2 & 3\\
            \hline
        \end{tabular}
        \par
    }
    \vspace{16pt}
    \caption{Processing and setup times of the tasks in \cref{illustrative_example}.}
    \label{example_times}
\end{table}

\subsection{Complexity of the Problem}
\label{sec:intro:complexity}
The problem addressed in this paper is a more general variant of the \NPHard\ problem $PD|seq, ser=1|C_{max}$ tackled in \emph{Vlk et al.}~\cite{vlk2019non}, where only one \cicmunda\ is considered and every task is dedicated to only one specific machine.

We can reduce instance of $PD|seq, ser=1|C_{max}$ to instance of $P|seq, ser|C_{max}$ by setting the setup times between all pairs of tasks dedicated to different machines in $PD|seq, ser=1|C_{max}$ instance to be infinity in $P|seq, ser|C_{max}$ instance. This makes the solution, where two tasks originally assigned to different machines are placed on the same machine infeasible. Since the availability of only one \cicmunda\ is implicitly supported by our problem definition, the reduction is complete, showing that $P|seq, ser|C_{max}$ is \NPHard.

\section{Related Work}
\label{sec:rel_work}
The existing related problems can be classified into three categories; first containing papers focused on sequence-dependent problems, second containing papers dedicated to problems with \cicmundove\ and third containing papers combining both properties. We discuss each in its section, while in the last one we also consider relations of existing problems to ours.

\subsection{Problems with Sequence-Dependent Setup Times}
\label{sec:rel_work:problems_without}
\emph{Allahverdi et al.}~\cite{allahverdi2008survey} survey showed that many papers tackling sequence-dependent setup scheduling exist. Papers like \emph{Vallada et al.}~\cite{vallada2011genetic} and \emph{Lee et al.}~\cite{lee1997scheduling} focused on heuristics, handling instances of up to lower hundreds of tasks in a few minutes of computational time. More recent papers usually focus on combination with other interesting properties.

\emph{Lunardi et al.}~\cite{LUNARDI2021419} focused on sequence-dependent scheduling with job decomposition, precedence constraints, preemption, unavailability periods and partial overlaps. Their metaheuristic was compared to CP proposed in \emph{Lunardi et al.}~\cite{LUNARDI2020105020}. Comparison indicated that CP alone could not handle some large instances and improved slower, but it guarantees improvement until an optimal solution is found.
This suggests that using a faster method to warm start the CP model, which further improves the solution, might be a promising way to tackle similar problems.

\emph{Rauchecker et al.}~\cite{RAUCHECKER2019338} tackled the problem of unrelated parallel machines with sequence-dependent setup times focusing on  high-performance computing and parallelization. It is shown that by applying their branch-and-price algorithm in a parallel manner, the computations can be reduced from hours to just minutes of computation.

\subsection{Problems with Servers}
\emph{Amir et al.}~\cite{amir2002scheduling} proposed an ILP formulation for a single \cicmunda, two parallel identical machines scheduling problem. They found that their model could not compute instances larger than 12 tasks. However, they show that when specific conditions are met, there exist fast approaches applicable in such constrained scenarios. A more effective block model ILP formulation for the two parallel identical machines problem was formulated by \emph{Hasani et al.}~\cite{HASANI201494}. This model provides an optimal solution to certain instances of up to 250 tasks in 3600 seconds. Alternatively the problem can be modeled by timed automata~\cite{WASZ}, but their performance is much lower.   

A scheduling and lot sizing problem with a common setup operator (\cicmunda) is studied in \emph{Tempelmeier et al.}~\cite{tempelmeier2008dynamic}. The setups performed by the operator are considered to be scheduled without overlaps. The setups are associated only with the following task, making them sequence-independent. The proposed ILP formulations can solve instances of tens of tasks, usually under a 1-minute time limit.

A problem involving setups performed by operators of different capabilities has been studied in \emph{Chen et al.}~\cite{chen2003optimization}. The problem is modelled using time-indexed formulation and solved by decomposition into smaller subproblems using Lagrangian relaxation. Subproblems are solved using Dynamic Programming (DP), and feasible solution is obtained by the composition of the subproblems. If that is impossible, the Lagrangian multipliers are updated using the surrogate subgradient method as in \emph{Zhao et al.}~\cite{zhao1999surrogate}. The downside is that the time-indexed formulation yields a model with pseudo-polynomial size, which is unsuitable if large processing and setup times are present. The proposed approach can handle instances of tens of machines in a matter of minutes.

\subsection{Problems with Sequence-Dependency and Servers}
\label{sec:rel_work:combined}
Papers \emph{Vlk et al.}~\cite{icores19} and \emph{Vlk et al.}~\cite{vlk2019non} study problem with sequence-dependent setups. However, tasks are dedicated to machines, and a single \cicmunda\ is assumed. CP models, an ILP model, and heuristics utilizing the problem's decomposition are proposed in both papers. The resulting subproblems deal with task ordering on machines independently. The results show that proposed CP models can find a solution for instances of tens of machines and tens of tasks for each machine in 1 minute. The proposed LOFAS algorithm with a heuristic can solve instances of up to 1000 tasks on 5 machines. The same problem as in \emph{Vlk et al.}~\cite{vlk2019non} is studied in \emph{Huang et al.}~\cite{hcz}. An ILP model and a Genetic Algorithm (GA) are proposed. It is stated that ILP formulation is unusable in a real-world scenario. The GA approach can solve instances of 10 machines and 100 tasks in under 50 seconds.

The problem in \emph{Kim et al.}~\cite{kl2} allows assignment of tasks to any machine. However, since setups are sequence-independent, there is no direct comparison in terms of generality to the aforementioned papers. Again, only one \cicmunda\ is allowed. ILP formulations and hybrid heuristics are proposed. The results show the ILP model can solve instances of 6 machines and up to 40 tasks, providing optimal or close to the optimal solution in 3600s. The hybrid heuristic solves the same instances under 1 minute, with resulting schedules \SIrange{2}{5}{\percent} longer than schedules obtained by ILP in 3600s.

Paper \emph{Hamzadayi et al.}~\cite{hy} generalizes \emph{Vlk et al.}~\cite{vlk2019non}, \emph{Huang et al.}~\cite{hcz} and \emph{Kim et al.}~\cite{kl2}, tackling machine-independent with sequence-dependent setups. However, it still considers only one \cicmunda, limiting its real-world applicability. ILP model is proposed, while the best performing proposed heuristic is GA. The results show that the ILP model can find the optimal solution in 18000 seconds for instances of 3 machines and 9 tasks. The GA can solve instances of 10 machines and 100 tasks in one minute.

Papers tackling more general problems than our considered $P|seq, ser|C_{max}$ also exist. In \emph{Costa et al.}~\cite{ijamt_2013}, a hybrid GA to a problem extended by \cicmunda's skill (\cicmunda's setup execution speed) and parallel unrelated machines is proposed. While the proposed GA can be applied to our problem, their experimental results show that its scalability is lower than the scalability of our proposed approach. Their proposed ILP model has no detailed results provided.

In more recent work by \emph{Yeppes-Borrero et al.}~\cite{YEPESBORRERO2020112959}, unrelated machines and setups requiring multiple \cicmundove\ are considered. Provided experimental results show that their proposed ILP model does not scale too well. While their proposed heuristic scales well, their results indicate it does not scale as well as our proposed approach.

\emph{Luis Fanjul-Peyro}~\cite{FANJULPEYRO2020100022} considers unrelated machines and multiple types of resources, i.e., \cicmundove. ILP model and exact approaches are proposed, but no heuristic approach is provided, so the scalability of the provided approaches is poor.

Other papers tackling more general problems like \emph{Lee et al.}~\cite{Lee2021}, \emph{Caricato et al.}~\cite{ijmt_2020}, \emph{Yeppes-Borrero et al.}~\cite{YEPESBORRERO2021443} or papers considering \cicmundove\ and sequence-dependency in different settings like \emph{Gnatowski et al.} \cite{9130513} exist, but their respective problem definitions are even more distant to ours than the problem definitions of the aforementioned papers.

In conclusion, research shows that while papers addressing less or more general problems exist, to the best of our knowledge, no paper tackles exactly our considered problem, which represents many real-world production scenarios. It is apparent from this section that as problem complexity increases, the ability to solve larger instances decreases rapidly, especially for exact approaches. While more general problem approaches do not provide the same efficiency and scalability as ours, our proposed approaches offer improved efficiency over existing approaches for less general problems. Furthermore, as seen in this section, the majority of works also employ ILP modelling as an exact approach. However, results from \emph{Vlk et al.}~\cite{vlk2019non} show that CP is far more suitable than ILP for this kind of scheduling problem. This contrasts with the current state of literature and is a good reason to investigate CP capability further.

\section{Constraint Programming Approach}
\label{sec:cp_exact}
The first approach we use to solve the considered $P|seq, ser|C_{max}$ problem is Constraint Programming (CP). Specifically, we use \emph{IBM ILOG CP Optimizer} CP solver.
The following few paragraphs describe basic CP concepts used. A reader familiar with CP formalism can skip to \cref{sec:cp_exact:model_formulation}, where model specifics are explained.

The main modelling expression used is \emph{interval variable}. As the name suggests, it represents some activity with its required reserved time in the schedule. Its length is denoted by \cpLengthOf, start time by \cpStartOf\ and completion time by \cpEndOf. While \cpLengthOf\ is given by the problem instance, \cpStartOf\ and \cpEndOf\ are determined by the CP solver. We use the interval variables to represent tasks and setups in our problem.

\begin{figure}[ht]
    \centering
    \captionsetup{justification=centering}
    \begin{tikzpicture}

\begin{scope}[xscale=1.0, yscale=0.8]

    \draw[dashed] (1.44,-0.75) -- (1.45,.25);
    \node[] at (1.5,-1) {\scriptsize \cpStartOf};
    \draw[dashed] (3.55,-0.75) -- (3.55,.25);
    \node[] at (3.5,-1) {\scriptsize \cpEndOf};
    
    \path[fill=lightgray] (0,0) rectangle (1,1);
    \draw (0,0) -- (1,0) -- (1,1) -- (0,1);
    
    \draw[fill=lightgray] (1.5, 0) rectangle node{\smaller \cpLengthOf} (3.5, 1);
    \draw[latex-latex, thick] (1.5, 0.2) -- (3.5, 0.2);
    
    \path[fill=lightgray] (4,0) rectangle (5,1);
    \draw (5,1) -- (4,1) -- (4,0) -- (5,0);
    
    \draw[thick] (0,-0.25) -- (5,-0.25);
    \draw[thick] (0,1.25) -- (5,1.25);
	
\end{scope}
	
\end{tikzpicture}
    \caption{Interval variable in the schedule.}
\end{figure}
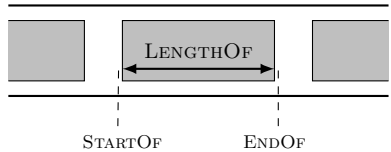

Every interval variable can be set \emph{optional}. Depending on the model's conditions and optimization criteria, the optional interval variable is \emph{present} or \emph{absent} in the solution. Unlike the present variable, the absent variable does not have to conform to the model's constraints. We can also specifically enforce the presence/absence of variables by introducing constraints implying either. One example is global constraint \cpAlternative\ which makes subset (usually of size one) of variables of given set present and rest absent while respecting other conditions in the model.

To model the relations we use \emph{constraints}. In addition to the linear algebraic constraints known from the ILP modelling, CP offers more complex ones such as logic constraints, scheduling constraints, global constraints and others. Global constraints are especially useful, enabling us to model specific predefined relations over a set of variables in a concise and computationally optimized way. One such constraint is the aforementioned \cpAlternative. This is a big reason why CP is so effective for some types of problems.

\subsection{Model Formulation}
\label{sec:cp_exact:model_formulation}
Since $PD|seq, ser=1|C_{max}$ problem considered in \emph{Vlk et al.}~\cite{vlk2019non} is a more constrained variant of our problem, we utilize some of the ideas proposed in it. There are two key ideas behind the proposed model. First, since \cicmundove\ are identical, we do not have to represent which specific \cicmunda\ is performing which setup. So rather than considering particular \cicmundove, we consider only how many \cicmundove\ we need at a particular time to perform the setups currently scheduled. Second, we do not represent every possible setup between a pair of tasks by a variable. Thus, we avoid the quadratic number of setup variables and related constraints; the number will equal the number of tasks. This is achieved by representing every possible setup following some particular task by one interval variable only.

When constructing the model, we must first ensure that every task is executed in the produced schedule. We denote the set of interval variables representing tasks as \taskIntervalSet\ and the interval variable of specific task \taskOne\ as \taskOneInterval. We use inequalities instead of equalities when assigning the processing times to variables in \taskIntervalSet\ so the interval variable representing task can prolong itself until a \cicmunda\ is available to perform the following setup. This will be useful later. Still, at least time equal to the task's processing time must be reserved in the schedule so that it can be executed. We assign the relaxed processing length to the interval variables as follows:
\setcounter{equation}{0}
\renewcommand{\theequation}{CP-\arabic{equation}}
\begin{equation}
    \label{cp_task_length}
    \forall {\taskOneIntervalEq \in \taskIntervalSetEq}: \cpLengthOf(\taskOneIntervalEq) \geq \taskOneProcessEq.
\end{equation}

Since we do not know which task will be assigned to which machine, we use the \cpAlternative\ constraint. For every \taskOneInterval\ $\in$ \taskIntervalSet\ and for every machine \machine\ $\in$ \machineSet\ we add a new optional interval variable \taskOneMachineInterval. Then for each task, all its optional variables are passed to the \cpAlternative\ constraint, ensuring that only one will be present. This way, every task will be present on exactly one machine. We also associate selected present \taskOneMachineInterval\ with the interval variable \taskOneInterval\ representing our task in other model's conditions by passing \taskOneInterval\ as an argument to \cpAlternative\ constraint. The resulting constraints are following:

\begin{equation}
    \label{cp_task_assignment}
    \forall {\taskOneIntervalEq \in \taskIntervalSetEq}:\,\cpAlternative \left(\taskOneIntervalEq, \bigcup_{{\machineEq \in \machineSetEq\\}} \left\{\taskOneMachineIntervalEq \right\}\right).
\end{equation}

To ensure that tasks do not overlap on any particular machine, for every machine, we add \cpNoOverlap\ constraint containing optional interval variables representing tasks scheduled on that machine. \cpNoOverlap\ constraint ensures that no two interval variables from a given set overlap in the produced schedule. Note that formally, interval variables passed to \cpNoOverlap\ constraint must be wrapped in sequence variable whose explanation we omit as it has no semantic significance in this case. \cpNoOverlap\ constraint can also be given \emph{transition matrix}, which defines empty interval sizes between the end of one and the start of the following interval variable in the given set. We use our setup times matrix \setupMatrix\ as the \emph{transition matrix} of the \cpNoOverlap\ constraint to insert setup times between tasks in the final machine schedules. The resulting formulation is following:
\begin{equation}
    \label{cp_task_nooverlap}
    \forall {\machineEq \in \machineSetEq}:\,\cpNoOverlap \left(\bigcup_{{\taskOneIntervalEq \in \taskIntervalSetEq\\}} \{\taskOneMachineIntervalEq \},\, \setupMatrixEq \right).
\end{equation}

We define a set \generalSetupIntervalSet\ of interval variables, where \generalSetupInterval\ represents setup following its respective \taskOneInterval\ from \taskIntervalSet. Notice that while setup times are already present between tasks thanks to \cref{cp_task_nooverlap}, we need these variables for \cicmunda\ constraints.
Since we do not need a setup after the last task on each machine, it will be a dummy setup of zero length. Due to that, we merely set the length of the setup variables to be at least zero, and the actual length of the setup executed will be determined later:
\begin{equation}
    \label{cp_setup_length}
    \forall {\generalSetupIntervalEq \in \generalSetupIntervalSetEq}: \cpLengthOf(\generalSetupIntervalEq) \geq 0.
\end{equation}

Next, we synchronize the start and completion times between tasks and setups. We use constraint \cpEndAtStart$(I_A, I_B)$, ensuring that the interval variable $I_A$ is completed exactly when the interval variable $I_B$ starts.
Using it, we can set the end of every task equal to the start of its following setup since the task's length is assigned by inequality and can be prolonged. The resulting constraints are following:
\begin{equation}
    \label{cp_task_setup_synchro}
    \forall {\taskOneIntervalEq \in \taskIntervalSetEq}: \cpEndAtStart(\taskOneIntervalEq, \generalSetupIntervalEq).
\end{equation}

\sloppy Now, we constrain ending times of all setups.
We achieve this by using \cpStartOfNext($I^{\prime}$) constraint, which returns the start time of the next interval variable following $I^{\prime}$ in the sequence. Thus, we can get the start time of the optional interval variable representing the next task after the given one in the machine sequence. Then, we set the end of the setup bigger or equal to the following task's start time. Using equality would not suffice because \cpStartOfNext\ evaluates to 0 for absent \taskOneMachineInterval\ and the last task on every machine. Note that even though the setup could theoretically be prolonged and end after the following task's start because of the inequality, it does not affect the feasibility of task placement and can be fixed by shortening such setups to the correct lengths after the solution is found. Thus, the final constraints are as follows:
\begin{equation}
    \label{cp_setup_end}
    \forall {\generalSetupIntervalEq \in \generalSetupIntervalSetEq}, \forall {\machineEq \in \machineSetEq}: \cpEndOf(\generalSetupIntervalEq) \ge \cpStartOfNext(\taskOneMachineIntervalEq).
\end{equation}

Finally, we set the maximum number of concurrently executed setups less or equal to the \cicmundove\ available. We achieve this by using the expression \cpPulse$(I^\prime, a)$. This expression specifies that $a$ units of resource, in our case \cicmundove, is used during interval $I^\prime$. The cumulative function is composed of \cpPulse\ terms for each \generalSetupInterval\ representing the use of precisely one \cicmunda. At any point in the schedule, the function is upper-bounded by the number of \cicmundove\ $\workerLastIndex$ provided:
\begin{equation}
    \label{cp_worker_pulse}
    \sum_{\generalSetupIntervalEq \in \generalSetupIntervalSetEq} \cpPulse(\generalSetupIntervalEq, 1) \leq \workerLastIndex.
\end{equation}

In the end, we minimize the makespan by minimizing the end of the last task:
\renewcommand{\theequation}{\arabic{equation}}
\renewcommand{\theequation}{CP-C$_{max}$}
\begin{equation}
    \label{cp_minimize_ends}
    \quad\textsc{Max}\left(\bigcup_{{\taskOneIntervalEq \in \taskIntervalSetEq}} \{\cpEndOf(\taskOneIntervalEq)\}\right).
\end{equation}

Thus, the complete CP model further denoted as \cp\ is defined as follows:
\renewcommand{\theequation}{\arabic{equation}}
\begin{equ}[!ht]
    \begin{empheq}[]{align}
               \textsc{Minimize}\quad & \eqref{cp_minimize_ends} \notag \\
             \text{s.t.}& \notag \\
            & \eqref{cp_task_length} - \eqref{cp_worker_pulse} \notag 
    \end{empheq}
\end{equ}
\setcounter{equation}{0}

An alternative CP model formulation where every setup was represented by its own interval variable was considered, but it was omitted from the paper due to having noticeably worse performance.

\section{Constructive Heuristic Algorithms}
\label{sec:constructive_heur}
Since the considered $P|seq, ser|C_{max}$ problem is \NPHard, using (only) an exact approach is computationally intractable for many larger problem instances. Therefore, it is essential to develop heuristic methods that produce a feasible solution of reasonable quality in an affordable time. In this section, two constructive heuristic algorithms are proposed:
\begin{itemize}
    \item \emph{Locally Optimal Selection of Setups} (\lss),
    \item \emph{Resolution of Setup Overlaps Lazily} (\rsl).
\end{itemize}

The main difference between the two is that \lss\ handles \cicmunda\ allocation during task scheduling while \rsl\ handles it in a separate phase after task sequences on machines are set. Three external functions called \textsc{GenerateStartingTasks}, \textsc{SelectNextTask} and \textsc{OptimizeScheduleEnds} are used by both algorithms:
\begin{itemize}
    \item \textsc{GenerateStartingTasks} selects a starting (first) task for every machine. The baseline version of this method selects tasks randomly.
    \item \textsc{SelectNextTask} provides a suitable task to follow after the current last one scheduled on a machine. The baseline version of this method selects the following task greedily by picking yet unscheduled task with the shortest setup time from the currently ending task to itself.
    \item \textsc{OptimizeScheduleEnds} re-optimizes the ending parts of machines' schedules after all tasks are assigned. This is important since the end of the machine schedule is the least compact and optimized part. The baseline version of this method only considers shortening the longest machine by moving the task to another machine.
\end{itemize}
Further improvements to these functions are proposed in \cref{sec:constructive_heur:improv}. For now, we consider them as black-box functions providing us with the described output.

To allow easy reproduction of the algorithms, described pseudocodes for both algorithms are provided in \ref{app:pseudocodes}.

\subsection{Locally Optimal Selection of Setups (LOSOS)}
\label{sec:constructive_heur:losos}
The idea of \lss\ is to iteratively construct chains of tasks on machines, always selecting the next task greedily according to some given evaluation function while at the same time considering \cicmunda\ availability.

The algorithm steps are following. First, \textsc{GenerateStartingTasks} is called to set starting tasks to machines. Then, a queue of \cicmundove\ is created, keeping track of times when \cicmundove\ are ready to work. Now, while there are tasks to schedule, the following is repeated. Machine with the currently shortest schedule is found. The task to follow the currently ending task is selected by \textsc{SelectNextTask}. Then, the soonest available \cicmunda\ is assigned to perform the setup. After all, tasks are assigned to machines, \textsc{OptimizeScheduleEnds} is called to re-optimize the schedule.

\begin{figure}[ht]
    \centering
    \captionsetup{justification=centering}

\scalebox{0.6}{
    \begin{tikzpicture}[
        node distance = 16mm and 12mm,
        start chain = A going right,
        base/.style = {draw, minimum width=32mm, minimum height=8mm, align=center, on chain=A},
        startstop/.style = {base, ellipse, rounded corners, fill=gray, minimum width=24mm},
        function/.style = {base, rectangle, fill=lightgray},
        io/.style = {base, trapezium, trapezium left angle=70, trapezium right angle=110, fill=lightgray!30},
        decision/.style = {base, diamond, aspect=2, fill=white, minimum width=24mm, minimum height=24mm},
        every edge quotes/.style = {auto=right}]
        \node [startstop]               {\LARGE Start};                                                    
        \node [function]                {\textsc{GenerateStartingTasks}};                           
        \node [function]                {\textsc{SelectNextTask}};                                  
        \node [io]                      {\large Assign soonest available \\ \large server to the setup};  
        \node [decision][below=of A-1]  {\large All tasks \\ scheduled};                                      
        \node [function]                {\textsc{OptimizeScheduleEnds}};                             
        \node [startstop]               {\LARGE Stop};                                                     
        
        
        \draw [arrows=-Stealth] 
            (A-1) edge[]    (A-2)
            (A-2) edge[]    (A-3)
            (A-3) edge[]    (A-4);
        \draw [arrows=-Stealth] 
            (A-4) -- +(0em, -4em)  node[anchor=east][below, xshift=-50mm]{} -|    (A-5);
        \draw [arrows=-Stealth] 
            (A-5.west) -+(-6.6em, -9.3em)-| +(0em, +12.8em)  node[anchor=east][above, xshift=+15mm]{no} -|       (A-3)
            (A-5) edge["yes"]   (A-6)
            (A-6) edge[]    (A-7)
            ;
    \end{tikzpicture}
}
    \caption{Flowchart of \lss\ algorithm.}
    \label{flowchart_losos}
\end{figure}
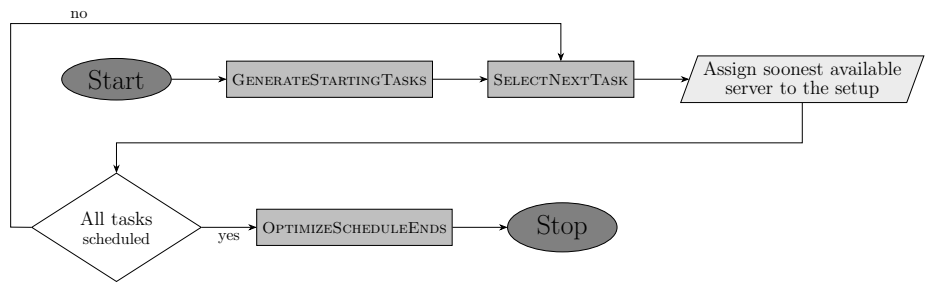

The algorithm is visualized in \cref{flowchart_losos}. It has polynomial time complexity  $\mathcal{O}((\log(\workerLastIndex) + \log(m) + t) \cdot t + m)$ if priority queues are used for \cicmundove. Baseline time complexities of \textsc{GenerateStartingTasks} and \textsc{OptimizeScheduleEnds} are considered to be $\mathcal{O}(m)$ and \textsc{SelectNextTask} baseline time complexity to be $\mathcal{O}(t)$.

\subsection{Resolution of Setup Overlaps Lazily (ROSOL)}
\label{sec:constructive_heur:rosol}
The problem of \lss\ is that it does not consider upcoming \cicmunda\ requirements when scheduling tasks and setups. To solve this, \rsl\ first ignores \cicmunda\ constraints when scheduling. Then, in the additional phase, further logic is applied to resolve these constraints violations in a more informed way.

The algorithm steps are following. First, \textsc{GenerateStartingTasks} is called to set starting tasks to machines.
Then, while there are unscheduled tasks, we repeatedly find the machine with the currently shortest schedule and assign task selected by \textsc{SelectNextTask} to it. When all tasks are scheduled, the problem is solved without \cicmundove' constraints, and \textsc{OptimizeScheduleEnds} is called to balance the schedule and reduce possible unnecessary conflicts arising when assigning the \cicmundove.

We ensure that \cicmundove' constraints hold in the following way. We move from schedule start to its end, checking if the number of concurrently performed setups $k$ does not exceed the number of \cicmundove\ $\workerLastIndex$. If it does, we need to postpone (move to the later time in the schedule) $k-\workerLastIndex$ setups until at least one \cicmunda\ performing non-postponed setup is free. The non-postponed setups are executed fully without preemption, meaning that they cannot be postponed once their execution started. We repeat this until we reach the end of the schedule. Finally, we call \textsc{OptimizeScheduleEnds} again.

The key remaining question is how to choose which setups to postpone. This is decided based on the \emph{tolerance coefficient} of the machine. Machine's \emph{tolerance coefficient} is equal to the difference between its schedule length and the current makespan plus the setup length currently to be executed on that machine. The \emph{tolerance coefficient} is calculated for every machine involved in the conflict, and setups on machines with $k-\workerLastIndex$ lowest tolerance coefficients are postponed.

\begin{figure}[ht]
    \centering
    \captionsetup{justification=centering}

\scalebox{0.6}{
    \begin{tikzpicture}[
        node distance = 20mm and 8mm,
        start chain = A going right,
        base/.style = {draw, minimum width=32mm, minimum height=8mm, align=center, on chain=A},
        startstop/.style = {base, ellipse, rounded corners, fill=gray, minimum width=24mm},
        function/.style = {base, rectangle, fill=lightgray},
        io/.style = {base, trapezium, trapezium left angle=70, trapezium right angle=110, fill=lightgray!30},
        decision/.style = {base, diamond, aspect=2, fill=white, minimum width=24mm, minimum height=24mm},
        every edge quotes/.style = {auto=right}]
        \node [startstop]               {\LARGE Start};                     
        \node [function]                {\textsc{GenerateStartingTasks}};   
        \node [function]                {\textsc{SelectNextTask}};          
        \node [decision]                {\large All tasks \\ scheduled};       
        \node [function]                {\textsc{OptimizeScheduleEnds}};     
        \node [io][below=of A-1]        {\large Move to next \\ \large possible setup conflict};    
        \node [decision]                {\large Conflict};                  
        \node [io][below=10mm and 10cm of A-7]        {\large Move $k-w$ setups forward}; 
        \node [decision][right=of A-7]  {\large Reached end \\ \large of schedule};                 
        \node [function]                {\textsc{OptimizeScheduleEnds}};     
        \node [startstop]               {\LARGE Stop};                      
        
        \draw [arrows=-Stealth] 
            (A-1) edge[]        (A-2)
            (A-2) edge[]        (A-3)
            (A-3) edge[]        (A-4)
            (A-4) edge["yes"]   (A-5);
        \draw [arrows=-Stealth] 
            (A-4) -- ($(A-4.south)!0.0!(A-3.south east)$) node[anchor=east][above, xshift=-20mm]{no}-|  (A-3);
        \draw [arrows=-Stealth] 
            (A-5) |- ($(A-5.south)!0.55!(A-6.north)$) -|       (A-6);
        \draw [arrows=-Stealth] 
            (A-6) edge[]        (A-7)
            (A-7) edge["yes"]   (A-8)
            (A-7) edge["no"]    (A-9)
            (A-8.west) -- +(+0em, -0em)  node[anchor=east][below, xshift=0mm]{} -|       (A-6)
            (A-9) -- +(+0em, -10em)  node[anchor=east][below, xshift=-10mm]{no} -|       (A-6)
            (A-9) edge["yes"]   (A-10)
            (A-10) edge[]       (A-11)
            ;
    \end{tikzpicture}
}
    \caption{Flowchart of \rsl\ algorithm.}
    \label{flowchart_rosol}
\end{figure}
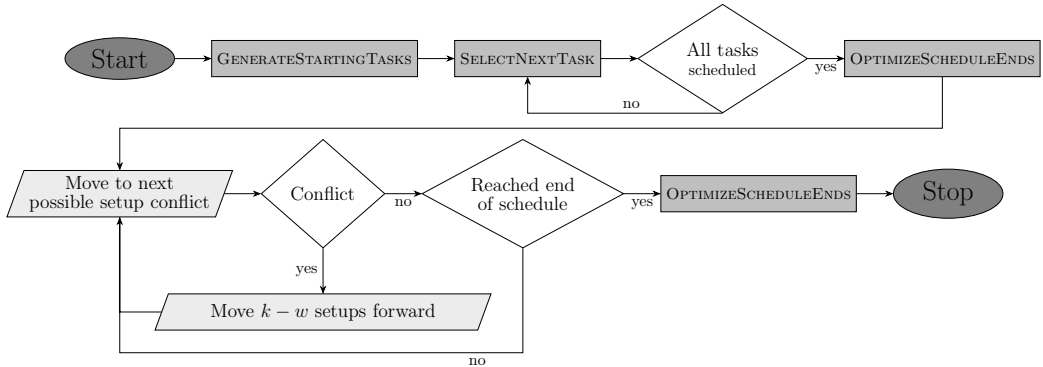

The algorithm is visualized in \cref{flowchart_rosol}. It has polynomial time complexity; $\mathcal{O}((m + \workerLastIndex \cdot \log(m) + t) \cdot t)$ if priority queues are used where possible and if only relevant places are checked for setup conflicts. Same as for \lss, we consider time complexities of external methods to be $\mathcal{O}(m)$ for \textsc{GenerateStartingTasks} and \textsc{OptimizeScheduleEnds} and $\mathcal{O}(t)$ for \textsc{SelectNextTask}.

\subsection{Constructive Heuristics Improvements}
\label{sec:constructive_heur:improv}
Algorithms \lss\ and \rsl\ are efficient in most cases, but they are greedy heuristics, exploring only a very limited part of solution space. For certain instances, this can lead to a solution very far from the optimum. Therefore, this section focuses on making both algorithms take less greedy and more informed decisions by extending methods \textsc{GenerateStartingTasks}, \textsc{OptimizeScheduleEnds} and \textsc{SelectNextTask}. The proposed improvements are:
\begin{itemize}
    \item Selecting starting tasks in an informed way. (\textsc{GenerateStartingTasks})
    \item Optimizing the endings of machine schedules with additional task swapping. (\textsc{OptimizeScheduleEnds})
    \item Improving setup/task selection using task priority coefficient. (\textsc{SelectNextTask})
    \item Improving setup/task selection using idleness reduction. (\textsc{SelectNextTask})
\end{itemize}

\subsubsection{Starting Tasks Selection}
\label{sec:constructive_heur:improv:start}
There is no way of choosing the best possible starting tasks without solving the entire problem, but we can devise methods that will, on average, improve the results. The description of the best performing method out of the tested ones follows.

Consider $T^{StartingTasks}$ denoting $m$ selected starting tasks and $z_j$ denoting $\min_{T_i \in T}(\setupTimeEq)$, i.e., the shortest setup from any $T_i \in T$ to a given $T_j$. Then, $T^{StartingTasks}$ are selected such that for every $T_{Starting} \in T^{StartingTasks}$, $z_{Starting}$ is one of the $m$ biggest $z_j$ from all $T_j \in T$. In other words, we are looking for $m$ tasks with their minimal preceding setup being one of $m$ biggest ones from all tasks. This way, we avoid performing setups for tasks that would require a long time, even when placed after their best compatible preceding task. Since only setup times vary depending on the sequence, testing shows a considerable improvement over picking tasks randomly. The asymptotic time complexity of the resulting method is $\mathcal{O}(t^2)$, thus having no impact on the asymptotic time complexity of \lss\ or \rsl\ when used.

Please note that this method works well because we assume all machines to execute their first task without a setup. If we would want this task selection method to work well in a case where setup is also necessary before the first task, we could add $m$ virtual zero-length tasks with required setup lengths to the other tasks and infinite setup lengths from all other tasks to these virtual tasks.

\subsubsection{End of Schedule Optimization}
\label{sec:constructive_heur:improv:end}
It has been observed that the most problematic part of the schedule is its end. The remaining tasks have only a few possible options where to be scheduled with their best compatible preceding tasks usually already unavailable, resulting in long setups and large differences between individual machine schedule lengths. Re-optimizing the end of the schedule reduces this effect. It is split into two phases:
\begin{enumerate}
    \item First, we find the machine with the longest schedule, denoted \machineLong. We take its last task and check if it can be moved to another machine while decreasing the makespan. If yes, we move it to the machine with the shortest resulting schedule after the move. We find new \machineLong\ and repeat this step until it decreases the makespan.
    \item Second, we switch ending tasks between pairs of machines to reduce the makespan further. If possible, a pair of machines is chosen so that the makespan decreases after the switch. If not, pair of machines is chosen such that the maximum schedule length of the pair decreases while the makespan does not increase, giving us a possibility for future improvement. The best pair in regards to reducing the makespan or machine schedule length is selected. Again, we repeat this until no machine pair meets the criteria.
\end{enumerate}

Potentially the optimization could be improved by looping the steps or considering more than just subsets of two machines. However, this would impact the time complexities of \lss\ and \rsl\, and the analysis of built schedules showed that the application of these two steps improved and equalized the schedule well enough. Considering the properties of machine schedules, the asymptotic time complexity of the resulting method is $\mathcal{O}(m^2)$, thus having no impact on the asymptotic time complexity of \lss\ or \rsl\ ($m \le t$) when used.

\subsubsection{Coefficient of Task Selection Priority}
\label{sec:constructive_heur:improv:coefficient}
The baseline version of \textsc{SelectNextTask} method used in \lss\ and \rsl\ chooses the next task to be scheduled greedily based on the setup length minimization. This works well for some distributions of setup lengths, but it has a fundamental flaw. As we deplete tasks with short setup times, it can happen that tasks having only long setups will remain, causing huge setbacks at the end of the schedule. The problem is further amplified because as long setups pile up at the end, they cause a shortage of available \cicmundove. If long setups were dispersed throughout the solution against the shorter ones, the problem would not occur. Hence, it is often better to take the locally suboptimal solution to obtain shorter setup times at the end of the schedule.

Thus, we are looking for a way of predicting which tasks should be saved for later and which are to be used immediately. As one possible way to evaluate this, we propose a function calculating so-called \emph{task coefficient} for the given task. We can think of this \emph{task coefficient} as a measure of future usefulness compared to the usefulness now. So unlike in the baseline version of \textsc{SelectNextTask}, where only the currently scheduled task and the next one to follow were considered, other possible future placements of the next task are considered as well.

The calculation is following. Let \taskOne\ represent currently ending scheduled task, \taskTwo\ represent task considered to follow \taskOne\ and $T_{unresolved}$ denote the set of tasks waiting to be scheduled or currently being executed. Then $T_{x1}, T_{x2}, T_{x3} \in T_{unresolved}$ denote tasks with smallest, second smallest and third smallest setup time to \taskTwo\ respectively, with setups denoted as $o_{x1,j}$, $o_{x2,j}$ and $o_{x3,j}$. The resulting \emph{task coefficient} is then calculated as follows:
\begin{dmath}
  \label{Coefficient}
  \emph{TaskCoefficient}_j(\taskOneEq, T_{unresolved}) =
  \underbrace{o_{i, j}^4 + }_{\text{current usefulness}}
  \underbrace{|o_{i, j} - o_{x1, j}| \cdot (o_{i, j} - o_{x1, j}) + |o_{i, j} - o_{x2, j}| \cdot (o_{i, j} - o_{x2, j}) + (o_{i, j} - o_{x3, j})}_{\text{possible future usefulness}}.
\end{dmath}

It is clear from \cref{Coefficient} that the largest emphasis is still on setup between \taskOne\ and \taskTwo, but other future scheduling possibilities are taken into account. To decide which task to follow after \taskOne, \emph{task coefficients} are calculated for all possible tasks to follow \taskOne, and the one with the smallest \emph{task coefficient} is chosen.

We considered several polynomial functions with different numbers and different powers of elements, but experiments showed that the one proposed in \cref{Coefficient} performs the best.
However, the efficiency of \emph{task coefficient} is influenced by the setup times distribution.
Its asymptotic time complexity is $\mathcal{O}(t^2)$ but since \textsc{SelectNextTask} is called $t$ times in both \lss\ and \rsl, it increases the overall complexity of both algorithms.

\subsubsection{Idleness Reduction}
\label{sec:constructive_heur:improv:setup}
We can further improve \textsc{SelectNextTask} method by considering future availabilities of \cicmundove\ when only one currently unoccupied one remains. This is only applicable when calling \textsc{SelectNextTask} from \lss\ as \rsl\ handles \cicmundove\ after task assignments and orders are already decided.

Let \taskOne\ denote the currently ending task, \taskTwo\ denote the next earliest ending task except for \taskOne\ and $T_{unscheduled}$ denote the set of yet unscheduled tasks. Let $\workerSet_l$ denote the only available unoccupied \cicmunda\ at the moment when the setup after \taskOne\ is being scheduled and $\workerSet_s$ denote the next earliest ending \cicmunda\ except for $\workerSet_l$. Now, let us consider two issues with the current way of setup scheduling:
\begin{enumerate}
    \item \label{set_over_red_1}
    If we choose $T_x \in T_{unscheduled}$ to follow after \taskOne\ with too long $o_{i,x}$ in-between, it might happen that when \taskTwo\ ends, both $\workerSet_l$ and $\workerSet_s$ will be occupied, thus no \cicmunda\ will be available to execute setup following \taskTwo. Therefore, the machine with $T_j$ scheduled on it would idle instead of executing the following setup.
    \item\label{set_over_red_2} If we assign too short $o_{i,x}$ to $\workerSet_l$, then after $o_{i,x}$ is finished, there might be no machine requiring the \cicmunda. In consequence, $\workerSet_l$ would idle instead of executing other longer $o_{i,x}$, which might have to be executed later, thus negatively affecting the makespan. Instead, executing a longer setup would maximize this free window's use and possibly shorten future setup times.
\end{enumerate}

To address issue \ref{set_over_red_1}, we calculate difference between ends of \taskTwo\ and \taskOne\ as $t_d = c_j - c_i$. Then, we choose task $T_x$ with $o_{i,x} \le t_d$, so after $\workerSet_l$ finishes setup between \taskOne\ and $T_x$, it will be ready to execute setup following \taskTwo\ once \taskTwo\ finishes. If more tasks have $o_{i,x} \le t_d$, we choose one according to the selection criteria, e.g. \emph{task coefficient}. In case there is no $T_x \in T_{unresolved}$ with $o_{i,x} \le t_d$, we pick $T_x$ with the smallest possible $o_{i,x}$ and again use selection criteria for possible tie breaking.

We only consider issue \ref{set_over_red_2} when we have some estimation of possible task usefulness elsewhere, like the aforementioned \emph{task coefficient}. Otherwise, we could potentially execute longer setups without a good reason. We calculate $t_d$ again but embed it into the \emph{task coefficient} calculation, changing $o_{i, j}^4$ in \emph{task coefficient} equation to $max(0, o_{i,j} - t_d)^4$. This way, we only emphasize the portion of the setup time, where the \cicmunda\ could already work on a different machine, weighing more other scheduling possibilities.

Addressing issue \ref{set_over_red_1} is more important than addressing issue \ref{set_over_red_2} since we rather let \cicmunda\ wait while machines execute as we minimize the makespan. Thus, if issue \ref{set_over_red_1} arises, we disregard issue \ref{set_over_red_2} and also any task utility metric like \emph{task coefficient}. It is clear that both issues and \emph{task coefficient} are partially in a trade-off, and some weighted decision between them could be derived, but this was not further examined. Since the \emph{task coefficient} must be executed because of the issue \ref{set_over_red_2} and the time complexity of additional operations is lower, the overall asymptotic time complexity is $\mathcal{O}(t^2)$. Since \textsc{SelectNextTask} is called $t$ times in both \lss\ and \rsl, it increases the overall complexity of both algorithms.

\subsection{Model Warm Starting: Synergy of the Approaches}
\label{sec:cp_exact:warm_start}
To improve the CP model's computational performance, we consider warm starting, a commonly used technique of obtaining a feasible or partial starting solution by using existing efficient algorithms for a similar problem. We developed and tested multiple warm starting methods but using the constructive heuristics proposed in this section yielded the best results as they provide quality solutions in a very short time.

Because the execution time of constructive heuristics is negligible compared to the CP model, we execute both and select the better solution as a warm start. Based on preliminary testing, the most efficient configuration is to execute \lss\ with all improvements and \rsl\ with starting task selection and end of schedule optimization. As we showed before, improvements to \textsc{SelectNextTask} method noticeably increase computational complexity, so executing \rsl\ without them saves computational time for the CP model. Also, since task coefficient and idleness reduction do not guarantee improvement, having diversity between \lss\ and \rsl\ solution is good.

The initial solution is passed to the CP model by assigning tasks to machines given the warm start solution assignment and ordering. We also add setup time spaces between tasks and set their start and end times accordingly.

\section{Experimental Evaluation}
\label{sec:experimental_evaluation}
In this section, we compare our proposed approaches to each other and then to the state-of-the-art approaches to similar problems. All experimental results were executed at a single core of the Intel Xeon 4110 processor running at 2.1GHz. A graphic card was not used in any of the calculations. Algorithms were implemented using C++. For CP, \emph{IBM ILOG CP Optimizer 20.1.0} was used, while for ILP, \emph{Gurobi 9.5} was used. All generated instances and solutions obtained by different approaches can be found enclosed on the GitLab repository \cite{Heinz2022}.

In all following tables, $m$ denotes the number of machines, $t$ the number of tasks and $\workerLastIndex$ the number of \cicmundove. By symbol $\infty$, we denote that no feasible solution was found. The approaches are compared by the Relative Difference (RD) calculated as $(compared - baseline) / baseline = RD$, where $compared$ denotes the objective value of the evaluated approach and $baseline$ denotes the objective value of the approach (or lower bound) to which the comparison is made.

\subsection{Comparison of Our Approaches}
\label{sec:experimental_evaluation:our_approaches}
We decided to compare all our approaches on the same instance set to show where their efficiencies meet. To make this possible, \machineLastIndex\ is set between 12 and 20, $t$ between 15\machineLastIndex\ and 20\machineLastIndex\ and $\workerLastIndex$\ is either 2 or 5. The processing and setup times of instances used in this subsection were randomly sampled from a uniform distribution between 1 and 50. The range of the distribution is based on \emph{Hall et al.}~\cite{Hall2001}\ which states that instances with both big and small variances of their attributes should be involved in the evaluation process. In subsection \ref{sec:experimental_evaluation:comparison}, we evaluate using methodologies given by the compared papers. Based on these comparisons, we show that our approaches are performant on various different ranges of attributes. However, no distribution of attributes in the compared papers has a range starting from 1. Thus, we include instances with such high variance in the comparison of our own. The uniform distribution was chosen as it is a common practice.

Since the constructive heuristics took between 1ms and 650ms to execute depending on the particular instance and improvements used, to obtain relevant comparison, we set CP model executions time limits considerably small. However, the results of supplementary experiments provided in \ref{app:cp_scalability} show that warm started CP models can be effectively applied to instances of up to 50 machines and 1000 tasks. With an hour time limit, the relative gap between the heuristics and CP model solutions is higher than on the tests performed in \cref{tab:our_approaches}.

In further text, \lss\ and \rsl\ denote algorithms without any improvements applied, \lsse\ denotes \lss\ execution with \textsc{GenerateStartingTasks} and \textsc{OptimizeScheduleEnds} improvements applied and \lsseic\ execution with all improvements applied. There is no \rsleic\ as idleness reduction cannot be applied to \rsl\, and the results for \rsle\ are not reported since the improvement achieved is very similar to \lsse. To distinguish CP model variants, \cp\ denotes model without warm start, \cpwse\ denotes model with better solution from \lsse\ and \rsle\ used as warm start and \cpwseic\ denotes model with better solution from \lsseic\ and \rsle\ used as a warm start. All results referenced in this subsection are in the \cref{tab:our_approaches}. The last column of the table contains lower bounds. The percentage in the round brackets in columns \lsseic\ and \cpwseic (10s) denotes RD to lower bound for each instance. The lower bound calculation with additional information can be found in \ref{app:lower_bound}.

\paragraph{Constructive heuristics} The difference between \lss\ and \rsl\ is not very pronounced, with \rsl\ being slightly better. A more noticeable difference between the two can be observed on instances with a smaller number of \cicmundove\ since \rsl\ assigns \cicmundove\ to setups in a more informed way. Regarding the effect of improvements proposed in \cref{sec:constructive_heur:improv}, based on the results from \cref{tab:our_approaches}, application of all improvements together yields more than \SI{4}{\percent} average decrease in the objective. We consider this enhancement notable given the sizes of relative gaps between the heuristics and the respective lower bounds provided in \cref{tab:our_approaches}.
However, for a few instances like number 1 or 14 in \cref{tab:our_approaches}, \lsseic\ does not yield the best solution out of all heuristics as it does not guarantee improvement. Thus, if constructive heuristics are used without the CP model, it would make sense to execute \lsse /\rsle\ and \lsseic\ and pick the better solution as \lsse\ has fairly negligible execution time. Regarding heuristics scalability, the graph provided in \ref{app:runtimes} shows that without the use of task coefficient and idleness reduction, even instances of hundreds of machines and tens of thousands of tasks can be solved in several seconds to several minutes of runtime.

\paragraph{CP models} Considering CP models, testing shows that \cp\ is not very usable for larger instances. The warm started variants yield incomparably better results than \cp\ in a fraction of its time limit. They also noticeably improve the warm started solution, showing that the model scales well and is effective once a reasonable starting solution is provided. In cases where the warm started model produced a solution with the same objective as the warm start, it was still able to construct the problem and search part of the solution space, meaning the problem was still feasible for the model in such a short time limit given. Using all improvements in warm starting showed beneficial as \cpwseic\ provided better results than \cpwse. Also, note that the warm starting is included within the time limit.

\begin{table}[H]
    \centering
    \scalebox{0.75}{
\setlength{\tabcolsep}{0.42em}
    \begin{tabular}{r|rrr|llll|lll|l}
    \hline
        \multicolumn{1}{c}{} & \multicolumn{3}{c}{\textbf{parameters}} & \multicolumn{8}{c}{{\textbf{objective value}} [-]}\\
       $\#$ & $m$ & $t$ & $r$ & \lss & \rsl & \lsse & \lsseic & \cp (120s) & \cpwse (10s) & \cpwseic (10s) & \fontfamily{bch}\selectfont{\small{LB}} \\
\hline
 1             & 12           & 180           & 2            & 436              & 418              & 427                 & 419 (9.7\%)            & 406                 & \textbf{398}           & 399 (4.5\%)               & 382                     \\
 2             & 12           & 180           & 5            & 418              & 418              & 418                 & 409 (7.1\%)            & \textbf{399}        & 403                    & 402 (5.2\%)               & 382                     \\
 3             & 12           & 240           & 2            & 580              & 563              & 553                 & 541 (8.4\%)            & 1798                & \textbf{515}           & 518 (3.8\%)               & 499                     \\
 4             & 12           & 240           & 5            & 545              & 545              & 543                 & 542 (8.6\%)            & 564                 & 519                    & \textbf{517 (3.6\%)}      & 499                     \\
 5             & 12           & 300           & 2            & 738              & 763              & 719                 & 704 (5.9\%)            & 7741                & 688                    & \textbf{685 (3.0\%)}      & 665                     \\
 6             & 12           & 300           & 5            & 743              & 743              & 712                 & 710 (6.8\%)            & 5492                & 686                    & \textbf{685 (3.0\%)}      & 665                     \\
 7             & 14           & 210           & 2            & 460              & 458              & 443                 & 430 (9.4\%)            & 1464                & 413                    & \textbf{412 (4.8\%)}      & 393                     \\
 8             & 14           & 210           & 5            & 440              & 440              & 434                 & 426 (8.4\%)            & \textbf{407}        & 415                    & 410 (4.3\%)               & 393                     \\
 9             & 14           & 280           & 2            & 612              & 615              & 595                 & 587 (8.7\%)            & 6406                & \textbf{563}           & \textbf{563 (4.3\%)}      & 540                     \\
 10            & 14           & 280           & 5            & 615              & 615              & 583                 & 569 (5.4\%)            & 6501                & 561                    & \textbf{557 (3.1\%)}      & 540                     \\
 11            & 14           & 350           & 2            & 759              & 741              & 752                 & 748 (8.7\%)            & 2950                & \textbf{727}           & \textbf{727 (5.7\%)}      & 688                     \\
 12            & 14           & 350           & 5            & 730              & 730              & 730                 & 720 (4.7\%)            & 3623                & 713                    & \textbf{710 (3.2\%)}      & 688                     \\
 13            & 16           & 240           & 2            & 456              & 414              & 433                 & 434 (16.0\%)           & 4823                & \textbf{395}           & \textbf{395 (5.6\%)}      & 374                     \\
 14            & 16           & 240           & 5            & 407              & 406              & 407                 & 408 (9.1\%)            & 1068                & \textbf{394}           & \textbf{394 (5.3\%)}      & 374                     \\
 15            & 16           & 320           & 2            & 603              & 609              & 591                 & 569 (6.8\%)            & 7775                & 573                    & \textbf{560 (5.1\%)}      & 533                     \\
 16            & 16           & 320           & 5            & 580              & 580              & 580                 & 557 (4.5\%)            & 8261                & 553                    & \textbf{551 (3.4\%)}      & 533                     \\
 17            & 16           & 400           & 2            & 732              & 728              & 708                 & \textbf{687 (5.7\%)}   & 9769                & 700                    & \textbf{687 (5.7\%)}      & 650                     \\
 18            & 16           & 400           & 5            & 707              & 707              & 698                 & \textbf{679 (4.5\%)}   & 6824                & 697                    & \textbf{679 (4.5\%)}      & 650                     \\
 19            & 18           & 270           & 2            & 436              & 433              & 411                 & 437 (19.7\%)           & 1717                & \textbf{400}           & 403 (10.4\%)              & 365                     \\
 20            & 18           & 270           & 5            & 411              & 411              & 402                 & 405 (11.0\%)           & 5758                & \textbf{382}           & \textbf{382 (4.7\%)}      & 365                     \\
 21            & 18           & 360           & 2            & 579              & 577              & 573                 & 552 (10.4\%)           & 7260                & \textbf{549}           & \textbf{549 (9.8\%)}      & 500                     \\
 22            & 18           & 360           & 5            & 557              & 550              & 539                 & 534 (6.8\%)            & 8522                & 533                    & \textbf{525 (5.0\%)}      & 500                     \\
 23            & 18           & 450           & 2            & 722              & 720              & 698                 & 703 (9.8\%)            & 7681                & \textbf{686}           & \textbf{686 (7.2\%)}      & 640                     \\
 24            & 18           & 450           & 5            & 717              & 717              & 686                 & \textbf{680 (6.2\%)}   & 8514                & 686                    & \textbf{680 (6.2\%)}      & 640                     \\
 25            & 20           & 300           & 2            & 493              & 472              & 473                 & 457 (14.5\%)           & 5630                & \textbf{445}           & \textbf{445 (11.5\%)}     & 399                     \\
 26            & 20           & 300           & 5            & 454              & 447              & 445                 & 434 (8.8\%)            & 7550                & 430                    & \textbf{419 (5.0\%)}      & 399                     \\
 27            & 20           & 400           & 2            & 588              & 577              & 568                 & 568 (9.4\%)            & 9822                & \textbf{547}           & \textbf{547 (5.4\%)}      & 519                     \\
 28            & 20           & 400           & 5            & 558              & 558              & \textbf{546}        & 551 (6.2\%)            & 8943                & \textbf{546}           & \textbf{546 (5.2\%)}      & 519                     \\
 29            & 20           & 500           & 2            & 765              & 752              & 743                 & 724 (10.9\%)           & 12811               & \textbf{705}           & \textbf{705 (8.0\%)}      & 653                     \\
 30            & 20           & 500           & 5            & 715              & 716              & 697                 & \textbf{676 (3.5\%)}   & 11874               & 695                    & \textbf{676 (3.5\%)}      & 653                     \\
 \hline
 $\sum$        & -            & -             & -            & 17556            & 17423            & 17107               & 16860                 & 172353              & 16517                  & \textbf{16414}           & 15600                   \\
 $RD$          & -            & -             & -            & 12.54 \%          & 11.69 \%          & 9.66 \%              & 8.08 \%                & 1004.83 \%           & 5.88 \%                 & \textbf{5.22 \%}          & 0.0 \%                   \\
 
\hline
\end{tabular}
}
    \caption{The comparison of all proposed approaches.}
    \label{tab:our_approaches}
\end{table}

\subsection{Comparison to Existing Approaches}
\label{sec:experimental_evaluation:comparison}
To evaluate whether our approaches have state-of-the-art quality, we compare them with existing ones. This is important as a comparison only between our approaches provides no external point of reference in regards to their overall quality. However, as the literature review revealed, there is no paper tackling our considered $P|seq, ser|C_{max}$ problem. Thus, we use problems $PD|seq, ser=1|C_{max}$, $P|ser=1|C_{max}$ and $P|seq, ser=1|C_{max}$ for the comparison as they are just restricted variants of $P|seq, ser|C_{max}$. The instances of said problems are transformed to the instances of $P|seq, ser|C_{max}$ so our approaches can be used to solve them. However, this makes the instances more complex than the original ones, even if their set of feasible solutions is the same, so our approaches might have to do more computations when solving them.

To compare the existing exact approaches realized by ILP models to our \cp, we formalized them using Gurobi 9.5. This way, we could make a direct comparison on the same hardware and with the same time limits, thus obtaining an entirely fair comparison. The problem instances used in the comparison were generated according to the descriptions provided in each respective paper.

As for the heuristic comparison, we compare our proposed heuristic alternative \cpwseic\ to existing heuristics. Since the comparison methodology differs for every paper, it is described in each paper's respective subsection. Again, all instances were generated according to their respective original paper descriptions.

All comparisons are made on instance sizes chosen by the original paper. Table layouts are not entirely consistent because they partially adopt layouts from their respective papers to make the comparison easier to evaluate.

\subsubsection{Dedicated Machines with Single Server} 
\label{sec:experimental_evaluation:comparison:HCZ}
In \emph{Huang et al.}~\cite{hcz}, problem denoted $PD|seq, ser=1|C_{max}$, where only 1 \cicmunda\ is present, and tasks are dedicated to machines is considered. The $PD|seq, ser=1|C_{max}$ instances were transformed to $P|seq, ser|C_{max}$ instances by setting infinite length setup times between tasks that are not supposed to be on the same machine. This way, if the CP solver would schedule tasks originally dedicated to different machines on the same machine, the resulting schedule would be infinitely long, thus infeasible.

There were no computational results provided for the MIP model in \emph{Huang et al.}~\cite{hcz}, only a statement that it performs poorly. In comparison, both ILP and \cp\ were given a 3600-second time limit. Even though \cp\ had to tackle larger problem instances, it performed much better, solving all instances and in most cases proving the optimality as well. The results can be seen in \cref{tab:hcz_1}. Note that $t / m$ in \cref{tab:hcz_1} denotes the number of tasks per machine.

\begin{table}[H]
    \centering
    \scalebox{0.75}{
    \begin{tabular}{l|llll|rr|rr}
        \toprule
        \multicolumn{1}{c}{} &\multicolumn{4}{c} {{\textbf{parameters}}} & \multicolumn{2}{c}{{\textbf{objective value [-]}}} & \multicolumn{2}{c}{{\textbf{CPU time [s]}}}\\
        $\#$ &   $m$ &   $t / m$ &   Setup LB &   Setup UB & \textsc{mip}\smaller{\cite{hcz}}   & \cp   & \textsc{mip}\smaller{\cite{hcz}}   & \cp   \\
        \hline
             1 &            2 &             4 &           5 &                  25 &              \textbf{329}   & \textbf{329}               & 0.154           & \textbf{0.015}             \\
             2 &            2 &             4 &           5 &                  50 &              \textbf{351}   & \textbf{351}               & 0.137           & \textbf{0.006}             \\
             3 &            2 &             4 &           25 &                  50 &              \textbf{394}   & \textbf{394}               & 0.109           & \textbf{0.014}             \\
             4 &            2 &             6 &           5 &                  25 &              \textbf{403}   & \textbf{403}               & 3.676           & \textbf{0.698}             \\
             5 &            2 &             6 &           5 &                  50 &              \textbf{423}   & \textbf{423}               & 22.444          & \textbf{0.082}             \\
             6 &            2 &             6 &           25 &                  50 &              \textbf{508}   & \textbf{508}               & 134.017         & \textbf{0.711}             \\
             7 &            2 &             8 &           5 &                  25 &              \textbf{469}   & \textbf{469}               & 682.419         & \textbf{1.194}             \\
             8 &            2 &             8 &           5 &                  50 &              \textbf{489}   & \textbf{489}               & 654.805         & \textbf{0.184}             \\
             9 &            2 &             8 &           25 &                  50 &              \textbf{627}   & \textbf{627}               & 3600.0          & \textbf{13.423}            \\
            10 &            3 &             6 &           5 &                  25 &              \textbf{419}   & \textbf{419}               & 54.212          & \textbf{1.443}             \\
            11 &            3 &             6 &           5 &                  50 &              \textbf{432}   & \textbf{432}               & 24.088          & \textbf{1.576}             \\
            12 &            3 &             6 &           25 &                  50 &              543            & \textbf{539}               & 3600.0          & \textbf{28.06}             \\
            13 &            3 &             9 &           5 &                  25 &              682            & \textbf{673}               & 3600.0          & \textbf{4.838}             \\
            14 &            3 &             9 &           5 &                  50 &              $\infty$         & \textbf{693}               & 3600.0          & \textbf{5.317}             \\
            15 &            3 &             9 &           25 &                  50 &              941            & \textbf{837}               & 3600.0          & \textbf{14.113}            \\
            16 &            3 &            12 &           5 &                  25 &              $\infty$         & \textbf{743}               & 3600.0          & \textbf{88.882}            \\
            17 &            3 &            12 &           5 &                  50 &              $\infty$         & \textbf{767}               & 3600.0          & \textbf{69.836}            \\
            18 &            3 &            12 &           25 &                  50 &              $\infty$         & \textbf{1042}              & \textbf{3600.0} & \textbf{3600.0}            \\
            19 &            4 &             8 &           5 &                  25 &              577            & \textbf{552}               & 3600.0          & \textbf{11.106}            \\
            20 &            4 &             8 &           5 &                  50 &              754            & \textbf{582}               & 3600.0          & \textbf{10.479}            \\
            21 &            4 &             8 &           25 &                  50 &              1063           & \textbf{831}               & \textbf{3600.0} & \textbf{3600.0}            \\
            22 &            4 &            12 &           5 &                  25 &              $\infty$         & \textbf{768}               & \textbf{3600.0} & \textbf{3600.0}            \\
            23 &            4 &            12 &           5 &                  50 &              $\infty$         & \textbf{813}               & \textbf{3600.0} & \textbf{3600.0}            \\
            24 &            4 &            12 &           25 &                  50 &              $\infty$         & \textbf{1278}              & \textbf{3600.0} & \textbf{3600.0}            \\
            25 &            4 &            16 &           5 &                  25 &              $\infty$         & \textbf{1078}              & 3600.0          & \textbf{2065.339}          \\
            26 &            4 &            16 &           5 &                  50 &              $\infty$         & \textbf{1097}              & \textbf{3600.0} & \textbf{3600.0}            \\
            27 &            4 &            16 &           25 &                  50 &              $\infty$         & \textbf{1678}              & \textbf{3600.0} & \textbf{3600.0}            \\
    \hline
    \end{tabular}
    }
    \caption{MIP model proposed in \emph{Huang et al.}~\cite{hcz} compared to \cp.}
    \label{tab:hcz_1}
\end{table}

The heuristic comparison in \emph{Huang et al.}~\cite{hcz} was realized by comparing their Genetic Algorithm (GA) to a lower bound denoted as LB$_2$. We calculated the LB$_2$ on the generated instances and compared RD between our CP model and LB$_2$ on said instances to their RD between GA and LB$_2$ on their instances (in paper denoted as GA-LB$_2$ gap), obtaining a reasonably fair comparison. Because \cpwseic\ uses constructive heuristics for warm starting and those do not respect machine dedication, we had to use much less effective \cp. The time limits were set according to the runtimes of the GA in the original paper. Note that $t / m$ in \cref{tab:hcz_2} denotes the number of tasks per machine.

The results in \cref{tab:hcz_2} show that out of 30 cases, \cp\ provided better RD to LB$_2$ than GA in 16 instances and worse in 13 instances. \cp\ excelled at solving smaller instances, but its effectivity decreased with increasing size. This is because we had to use \cp\ instead of \cpwseic, solved a larger problem instance than the GA and also because of the very short time limit, which would generally favour heuristics. The largest tested instance had only a 30-second time limit. In conclusion, the comparison shows that even when tackling a much more restricted case of our problem with our less tailored approach without its main improvements applied, it still produces competitive results.

\begin{table}[H]
    \centering
    \scalebox{0.75}{
    \begin{tabular}{l|llll|rr|rr}
        \toprule
        \multicolumn{1}{c}{} &\multicolumn{4}{c} {{\textbf{parameters}}} & \multicolumn{2}{c}{\textbf{objective value [-]}} & \multicolumn{2}{c}{\textbf{RD [\%]}}\\
        $\#$ & $m$ &   $t / m$ &   Setup LB &   Setup UB &   LB2 &   \cp & GA\smaller{\cite{hcz}}   & \cp   \\
        \hline
             1 &            2 &            5 &                   5 &                  25 &            380 &                  380 & \textbf{0.0}   & \textbf{0.0}         \\
             2 &            2 &            5 &                   5 &                  50 &            390 &                  390 & 0.47           & \textbf{0.0}         \\
             3 &            2 &            5 &                  25 &                  50 &            463 &                  463 & 0.43           & \textbf{0.0}         \\
             4 &            2 &            10 &                   5 &                  25 &            671 &                  671 & 0.46           & \textbf{0.0}         \\
             5 &            2 &            10 &                   5 &                  50 &            691 &                  691 & 1.1            & \textbf{0.0}         \\
             6 &            2 &            10 &                  25 &                  50 &            855 &                  855 & 1.08           & \textbf{0.0}         \\
             7 &            3 &            5 &                   5 &                  25 &            380 &                  380 & 0.04           & \textbf{0.0}         \\
             8 &            3 &            5 &                   5 &                  50 &            390 &                  390 & 0.42           & \textbf{0.0}         \\
             9 &            3 &            5 &                  25 &                  50 &            463 &                  463 & 1.56           & \textbf{0.0}         \\
            10 &            3 &            10 &                   5 &                  25 &            689 &                  689 & 0.57           & \textbf{0.0}         \\
            11 &            3 &            10 &                   5 &                  50 &            717 &                  717 & 1.51           & \textbf{0.0}         \\
            12 &            3 &            10 &                  25 &                  50 &            874 &                  926 & \textbf{1.82}  & 5.95                 \\
            13 &            5 &            5 &                   5 &                  25 &            380 &                  380 & 0.09           & \textbf{0.0}         \\
            14 &            5 &            5 &                   5 &                  50 &            390 &                  390 & 1.94           & \textbf{0.0}         \\
            15 &            5 &            5 &                  25 &                  50 &            552 &                  605 & \textbf{2.03}  & 9.6                  \\
            16 &            5 &            10 &                   5 &                  25 &            689 &                  689 & 0.69           & \textbf{0.0}         \\
            17 &            5 &            10 &                   5 &                  50 &            717 &                  740 & \textbf{2.14}  & 3.21                 \\
            18 &            5 &            10 &                  25 &                  50 &           1225 &                 1353 & \textbf{5.48}  & 10.45                \\
            19 &            7 &            5 &                   5 &                  25 &            380 &                  380 & 0.48           & \textbf{0.0}         \\
            20 &            7 &            5 &                   5 &                  50 &            390 &                  413 & \textbf{4.38}  & 5.9                  \\
            21 &            7 &            5 &                  25 &                  50 &            769 &                  827 & \textbf{1.89}  & 7.54                 \\
            22 &            7 &            10 &                   5 &                  25 &            689 &                  692 & 1.18           & \textbf{0.44}        \\
            23 &            7 &            10 &                   5 &                  50 &            717 &                  846 & \textbf{6.37}  & 17.99                \\
            24 &            7 &            10 &                  25 &                  50 &           1683 &                 1823 & \textbf{5.94}  & 8.32                 \\
            25 &           10 &            5 &                   5 &                  25 &            402 &                  402 & 2.36           & \textbf{0.0}         \\
            26 &           10 &            5 &                   5 &                  50 &            418 &                  541 & \textbf{6.99}  & 29.43                \\
            27 &           10 &            5 &                  25 &                  50 &           1112 &                 1180 & \textbf{2.31}  & 6.12                 \\
            28 &           10 &           10 &                   5 &                  25 &            689 &                  800 & \textbf{3.81}  & 16.11                \\
            29 &           10 &           10 &                   5 &                  50 &            717 &                 1186 & \textbf{22.98} & 65.41                \\
            30 &           10 &           10 &                  25 &                  50 &           2391 &                 2568 & \textbf{6.87}  & 7.4                  \\
        \hline
    \end{tabular}
    }
    \caption{GA proposed in \emph{Huang et al.}~\cite{hcz} compared to \cp.}
    \label{tab:hcz_2}
\end{table}

\subsubsection{Sequence-Independent Setups with Single Server}
\label{sec:experimental_evaluation:comparison:KL2}
In \emph{Kim et al.}~\cite{kl2}, problem denoted $P|ser=1|C_{max}$, where only 1 \cicmunda\ is present, and setups are sequence-independent and executed before every task is considered. The $P|ser=1|C_{max}$ instances were transformed to $P|seq, ser|C_{max}$ instances in two steps. First, the instance's setup times matrix was extended to sequence-dependent format by simply duplicating the setup times. Second, $m$ zero-length tasks were added to the instance, each one ending up as the first task on one of the machines. Then, the setups between these zero-length tasks and their following real tasks emulate the setup before the first task on each machine in $P|ser=1|C_{max}$ problem.

We implemented a better performing MIP model from \emph{Kim et al.}~\cite{kl2} denoted as \textsc{mip-2} and compared it to our \cp. \cref{tab:kl2_1} shows the comparison results for instances of 6 machines with either 20, 30, or 40 tasks, with two additional settings, $\alpha$ and $p$. Parameter $\alpha$ represents the amount of variance in task and setup times, where 0.1 means a maximum of \SI{10}{\percent} deviation from average on either side. Parameter $p$ describes a multiplier of the setup length; the bigger it is, the longer the setups can be. The results show that \cp\ always returned the same or better solution than the MIP model. The models were given a 3600-second time limit.
\begin{table}[H]
    \centering
    \scalebox{0.75}{
    \begin{tabular}{ll|rr|rr|rr}
        \toprule
        \multicolumn{2}{r}{{\textbf{parameters}}} & \multicolumn{2}{c}{\textbf{(20, 6)}}& \multicolumn{2}{c}{\textbf{(30, 6)}}& \multicolumn{2}{c}{\textbf{(40, 6)}}\\
        {$\alpha$} &   $p$ & \textsc{mip-2}\smaller{\cite{kl2}}   & \cp   &   \textsc{mip-2}\smaller{\cite{kl2}} & \cp   &   \textsc{mip-2}\smaller{\cite{kl2}} & \cp   \\
        \hline
              0.1 &          0.5 & \textbf{205}     & \textbf{205}               &              284 & \textbf{278}               &              382 & \textbf{370}               \\
              0.1 &          0.7 & 213              & \textbf{210}               &              291 & \textbf{288}               &              390 & \textbf{380}               \\
              0.1 &          1   & \textbf{225}     & \textbf{225}               &              322 & \textbf{316}               &              420 & \textbf{413}               \\
              0.3 &          0.5 & 190              & \textbf{186}               &              278 & \textbf{274}               &              375 & \textbf{366}               \\
              0.3 &          0.7 & 193              & \textbf{192}               &              291 & \textbf{282}               &              383 & \textbf{375}               \\
              0.3 &          1   & \textbf{212}     & \textbf{212}               &              314 & \textbf{307}               &              415 & \textbf{403}               \\
              0.5 &          0.5 & 190              & \textbf{186}               &              285 & \textbf{279}               &              381 & \textbf{373}               \\
              0.5 &          0.7 & 193              & \textbf{189}               &              293 & \textbf{282}               &              387 & \textbf{377}               \\
              0.5 &          1   & 213              & \textbf{211}               &              324 & \textbf{309}               &              421 & \textbf{408}               \\
        \hline
    \end{tabular}
    }
    \caption{MIP-2 model proposed in \emph{Kim et al.}~\cite{kl2} compared to \cp.}
    \label{tab:kl2_1}
\end{table}

To evaluate the efficiency of the hybrid heuristic denoted as HA in \emph{Kim et al.}~\cite{kl2}, authors calculated RD (in paper denoted as gap percentage) between HA results and results of \textsc{mip-2} running for 3600 seconds. Because we obtained \textsc{mip-2} results in the previous comparison, we can compare HA against \cpwseic\ by comparing their calculated RD with RD between \cpwseic\ and \textsc{mip-2}. This comparison should be fair even if there is any difference between their and our hardware because the \textsc{mip-2} results and the \cpwseic\ results were obtained on the same computer. Thus, if our machine would be faster, the \textsc{mip-2} results to which \cpwseic\ is compared would also profit from the computational power increase. \cpwseic's time limit was set according to the HA's runtime limit stated in \emph{Kim et al.}~\cite{kl2}.

The results in \cref{tab:kl2_2} show that for every single instance, \cpwseic\ provided a better solution than HA. In 25 out of 30 instances, \cpwseic\ with limited time also provided better results than \textsc{mip-2} running for 3600 seconds, with only 2 instances being worse. We believe that the consistency and size of the difference ensure that our approach improves over the compared one, even if the comparison method would slightly influence the results.

\begin{table}[H]
    \centering
    \scalebox{0.75}{
    \begin{tabular}{l|llll|rr|rr}
        \toprule
        \multicolumn{1}{c}{} &\multicolumn{4}{c}{{\textbf{parameters}}} & \multicolumn{2}{c}{{\textbf{objective value [-]}}}& \multicolumn{2}{c}{{\textbf{RD [\%]}}}\\
        $\#$ & $m$ &   $t$ &   $p$ &   \textbf{$\alpha$} &   \textsc{mip-2}\smaller{\cite{kl2}} &   \cpwseic &   HA\smaller{\cite{kl2}}/\larger{\textsc{mip-2}}\smaller{\cite{kl2}} & \cpwseic/\textsc{mip-2}\smaller{\cite{kl2}}   \\
        \hline
             1 &            6 &            20 &          0.5 &              0.1 &            205 &               205 &               1.87 & \textbf{0.0}       \\
             2 &            6 &            20 &          0.7 &              0.1 &            213 &               210 &               2.34 & \textbf{-1.41}     \\
             3 &            6 &            20 &          1   &              0.1 &            225 &               225 &               2.72 & \textbf{0.0}       \\
             4 &            6 &            20 &          0.5 &              0.3 &            190 &               186 &               1.06 & \textbf{-2.11}     \\
             5 &            6 &            20 &          0.7 &              0.3 &            193 &               193 &               2.15 & \textbf{0.0}       \\
             6 &            6 &            20 &          1   &              0.3 &            212 &               216 &               2.72 & \textbf{1.89}      \\
             7 &            6 &            20 &          0.5 &              0.5 &            190 &               187 &               3.15 & \textbf{-1.58}     \\
             8 &            6 &            20 &          0.7 &              0.5 &            193 &               190 &               2.02 & \textbf{-1.55}     \\
             9 &            6 &            20 &          1   &              0.5 &            213 &               214 &               4.03 & \textbf{0.47}      \\
            10 &            6 &            30 &          0.5 &              0.1 &            284 &               280 &               1.24 & \textbf{-1.41}     \\
            11 &            6 &            30 &          0.7 &              0.1 &            291 &               289 &               2.21 & \textbf{-0.69}     \\
            12 &            6 &            30 &          1   &              0.1 &            322 &               320 &               1.74 & \textbf{-0.62}     \\
            13 &            6 &            30 &          0.5 &              0.3 &            278 &               275 &               1.29 & \textbf{-1.08}     \\
            14 &            6 &            30 &          0.7 &              0.3 &            291 &               283 &               2.65 & \textbf{-2.75}     \\
            15 &            6 &            30 &          1   &              0.3 &            314 &               313 &               2.42 & \textbf{-0.32}     \\
            16 &            6 &            30 &          0.5 &              0.5 &            285 &               279 &               2.68 & \textbf{-2.11}     \\
            17 &            6 &            30 &          0.7 &              0.5 &            293 &               284 &               2.9  & \textbf{-3.07}     \\
            18 &            6 &            30 &          1   &              0.5 &            324 &               312 &               2.7  & \textbf{-3.7}      \\
            19 &            6 &            40 &          0.5 &              0.1 &            382 &               372 &               2.22 & \textbf{-2.62}     \\
            20 &            6 &            40 &          0.7 &              0.1 &            390 &               382 &               3.55 & \textbf{-2.05}     \\
            21 &            6 &            40 &          1   &              0.1 &            420 &               415 &               2.05 & \textbf{-1.19}     \\
            22 &            6 &            40 &          0.5 &              0.3 &            375 &               367 &               1.4  & \textbf{-2.13}     \\
            23 &            6 &            40 &          0.7 &              0.3 &            383 &               376 &               2.7  & \textbf{-1.83}     \\
            24 &            6 &            40 &          1   &              0.3 &            415 &               407 &               4.66 & \textbf{-1.93}     \\
            25 &            6 &            40 &          0.5 &              0.5 &            381 &               374 &               0.91 & \textbf{-1.84}     \\
            26 &            6 &            40 &          0.7 &              0.5 &            387 &               378 &               2.36 & \textbf{-2.33}     \\
            27 &            6 &            40 &          1   &              0.5 &            421 &               414 &               3.86 & \textbf{-1.66}     \\
\hline
    \end{tabular}
    }
    \caption{HA proposed in \emph{Kim et al.}~\cite{kl2} compared to \cpwseic.}
    \label{tab:kl2_2}
\end{table}

\subsubsection{Single Server}
\label{sec:experimental_evaluation:comparison:HY}
In \emph{Hamzadayi et al.}~\cite{hy}, $P|seq, ser=1|C_{max}$ problem is considered, meaning the only difference compared to our problem is the presence of just 1 \cicmunda. For the exact approach comparison, we changed the original testing time limit from 18000 to 3600 seconds to obtain the MIP model and \cp\ results in a more reasonable time. It can be seen from the \cref{hy_1} that \cp\ was always on par or better than the MIP model. In fact, in cases where results were the same, we know that \cp\ reached the optimal solution. Model \cp\ was also able to prove optimality in some instances where the MIP model did not even reach the optimal solution.
\noindent
\begin{table}[H]
    \centering
    \scalebox{0.75}{
\begin{tabular}{l|ll|rr|rr}
    \toprule
    \multicolumn{1}{c}{} & \multicolumn{2}{c}{{\textbf{parameters}}} & \multicolumn{2}{c}{{\textbf{objective value [-]}}}& \multicolumn{2}{c}{{\textbf{CPU time [s]}}}\\
    $\#$ &   $m$ &   $t$ & \textsc{mip}\smaller{\cite{hy}}   & \cp   & \textsc{mip}\smaller{\cite{hy}}   & \cp   \\
    \hline
             1 &            2 &             6 & \textbf{200}   & \textbf{200}               & 2.135           & \textbf{0.043}             \\
             2 &            2 &             8 & \textbf{237}   & \textbf{237}               & 144.502         & \textbf{0.305}             \\
             3 &            2 &            10 & \textbf{334}   & \textbf{334}               & 3600.0          & \textbf{0.957}             \\
             4 &            2 &            14 & 441            & \textbf{415}               & 3600.0          & \textbf{3600.0}          \\
             5 &            2 &            20 & 687            & \textbf{611}               & 3600.0          & \textbf{3600.0}          \\
             6 &            3 &             9 & \textbf{188}   & \textbf{188}               & 577.058         & \textbf{0.696}             \\
             7 &            3 &            12 & 279            & \textbf{253}               & 3600.0          & \textbf{7.166}             \\
             8 &            3 &            15 & 376            & \textbf{346}               & 3600.0          & \textbf{3600.0}          \\
             9 &            3 &            21 & $\infty$         & \textbf{373}               & \textbf{3600.0} & \textbf{3600.0}            \\
            10 &            3 &            30 & $\infty$         & \textbf{667}               & \textbf{3600.0} & \textbf{3600.0}            \\
            11 &            4 &            12 & 205            & \textbf{196}               & 3600.0          & \textbf{15.741}            \\
            12 &            4 &            16 & 231            & \textbf{199}               & 3600.0          & \textbf{3600.0}          \\
            13 &            4 &            20 & 381            & \textbf{309}               & \textbf{3600.0} & \textbf{3600.0}            \\
            14 &            4 &            28 & $\infty$         & \textbf{448}               & \textbf{3600.0} & \textbf{3600.0}            \\
            15 &            4 &            40 & $\infty$         & \textbf{656}               & \textbf{3600.0} & \textbf{3600.0}            \\
            16 &            5 &            15 & 230            & \textbf{209}               & 3600.0          & \textbf{1972.226}          \\
            17 &            5 &            20 & 369            & \textbf{247}               & \textbf{3600.0} & \textbf{3600.0}            \\
            18 &            5 &            25 & 668            & \textbf{299}               & \textbf{3600.0} & \textbf{3600.0}            \\
            19 &            5 &            35 & $\infty$         & \textbf{458}               & \textbf{3600.0} & \textbf{3600.0}            \\
            20 &            5 &            50 & $\infty$         & \textbf{669}               & \textbf{3600.0} & \textbf{3600.0}            \\
            21 &            7 &            21 & 265            & \textbf{167}               & 3600.0          & \textbf{3600.0}          \\
            22 &            7 &            28 & $\infty$         & \textbf{264}               & \textbf{3600.0} & \textbf{3600.0}            \\
            23 &            7 &            35 & $\infty$         & \textbf{334}               & \textbf{3600.0} & \textbf{3600.0}            \\
            24 &            7 &            49 & $\infty$         & \textbf{425}               & \textbf{3600.0} & \textbf{3600.0}            \\
            25 &            7 &            70 & $\infty$         & \textbf{640}               & \textbf{3600.0} & \textbf{3600.0}            \\
            26 &           10 &            30 & $\infty$         & \textbf{210}               & \textbf{3600.0} & \textbf{3600.0}            \\
            27 &           10 &            40 & $\infty$         & \textbf{275}               & \textbf{3600.0} & \textbf{3600.0}            \\
            28 &           10 &            50 & $\infty$         & \textbf{355}               & \textbf{3600.0} & \textbf{3600.0}            \\
            29 &           10 &            70 & $\infty$         & \textbf{457}               & \textbf{3600.0} & \textbf{3600.0}            \\
            30 &           10 &           100 & $\infty$         & \textbf{678}               & \textbf{3600.0} & \textbf{3600.0}            \\
\hline
\end{tabular}
}
    \caption{MIP model proposed in \emph{Hamzadayi et al.}~\cite{hy} compared to \cp.}
    \label{hy_1}
\end{table}

The authors in \emph{Hamzadayi et al.}~\cite{hy} evaluated their GA effectivity by comparing it to their other proposed approaches. However, most instances had no results for the MIP model, so we had no way of fairly comparing \cpwseic\ to the GA without executing the GA ourselves.
Thus, we reimplemented the GA according to the paper's description and to the best of our knowledge, denoting it in the following text as GA$_R$.

The results in \cref{tab:hy_2} show \cpwseic\ outperforming GA$_R$ in every case except two, where they yielded the same result. The runtime of \cpwseic\ for each instance was set equal to the runtime obtained by GA$_R$ on our hardware, which was lower than runtimes reported in \emph{Hamzadayi et al.}~\cite{hy}. Thus, \cpwseic\ had no unfair advantage considering the time limit. Every machine-task combination was run 20 times to even out the GA's random nature, and the rounded average was reported. The last column shows the RD of objective values of GA$_R$ to \cpwseic. It is noticeable that the difference gets more significant with the growing instance's size. This is probably the result of GA's inability to effectively find improvements in such a vast solution space.
\noindent
\begin{table}[H]
    \centering
    \scalebox{0.75}{
\begin{tabular}{l|ll|rr|r}
    \toprule
    \multicolumn{1}{c}{} & \multicolumn{2}{c}{{\textbf{parameters}}} & \multicolumn{2}{c}{{\textbf{objective value [-]}}} & \multicolumn{1}{c}{\textbf{{RD [\%]}}}\\
    $\#$ &   $m$ &   $t$ & \cpwseic   & GA$_R$\smaller{\cite{hy}}   &   GA$_R$\smaller{\cite{hy}} to \cpwseic \\
\hline  
             1 &            2 &             6 & \textbf{201}     & \textbf{201}         & 0.0 \%                      \\
             2 &            2 &             8 & \textbf{255}     & 256                  & 0.39 \%                     \\
             3 &            2 &            10 & \textbf{333}     & 337                  & 1.2 \%                      \\
             4 &            2 &            14 & \textbf{443}     & 464                  & 4.74 \%                     \\
             5 &            2 &            20 & \textbf{650}     & 693                  & 6.62 \%                     \\
             6 &            3 &             9 & \textbf{200}     & \textbf{200}         & 0.0 \%                      \\
             7 &            3 &            12 & \textbf{267}     & 271                  & 1.5 \%                      \\
             8 &            3 &            15 & \textbf{331}     & 340                  & 2.72 \%                     \\
             9 &            3 &            21 & \textbf{451}     & 479                  & 6.21 \%                     \\
            10 &            3 &            30 & \textbf{643}     & 696                  & 8.24 \%                     \\
            11 &            4 &            12 & \textbf{204}     & 211                  & 3.43 \%                     \\
            12 &            4 &            16 & \textbf{253}     & 262                  & 3.56 \%                     \\
            13 &            4 &            20 & \textbf{332}     & 351                  & 5.72 \%                     \\
            14 &            4 &            28 & \textbf{458}     & 495                  & 8.08 \%                     \\
            15 &            4 &            40 & \textbf{649}     & 714                  & 10.02 \%                    \\
            16 &            5 &            15 & \textbf{206}     & 213                  & 3.4 \%                      \\
            17 &            5 &            20 & \textbf{272}     & 287                  & 5.51 \%                     \\
            18 &            5 &            25 & \textbf{332}     & 355                  & 6.93 \%                     \\
            19 &            5 &            35 & \textbf{447}     & 492                  & 10.07 \%                    \\
            20 &            5 &            50 & \textbf{661}     & 741                  & 12.1 \%                     \\
            21 &            7 &            21 & \textbf{212}     & 223                  & 5.19 \%                     \\
            22 &            7 &            28 & \textbf{280}     & 304                  & 8.57 \%                     \\
            23 &            7 &            35 & \textbf{334}     & 373                  & 11.68 \%                    \\
            24 &            7 &            49 & \textbf{460}     & 528                  & 14.78 \%                    \\
            25 &            7 &            70 & \textbf{646}     & 755                  & 16.87 \%                    \\
            26 &           10 &            30 & \textbf{230}     & 251                  & 9.13 \%                     \\
            27 &           10 &            40 & \textbf{294}     & 327                  & 11.22 \%                    \\
            28 &           10 &            50 & \textbf{369}     & 421                  & 14.09 \%                    \\
            29 &           10 &            70 & \textbf{495}     & 580                  & 17.17 \%                    \\
            30 &           10 &           100 & \textbf{713}     & 844                  & 18.37 \%                    \\
\hline
\end{tabular}
}
    \caption{GA proposed in \emph{Hamzadayi et al.}~\cite{hy} compared to \cpwseic.}
    \label{tab:hy_2}
\end{table}

\section{Conclusion and Future Work}
\label{sec:conclusion}
This paper addressed the scheduling of tasks on non-dedicated machines with sequence-dependent setups that are performed by \cicmundove. Initially, the Constraint Programming model was proposed to formalize and optimally solve the problem. However, the CP model is suitable only for specific instance sizes. Thus, two polynomial constructive heuristics \lss\ and \rsl\ were proposed. Since the execution of constructive heuristics is very fast, improvements to these methods were proposed, enhancing the solution's quality in a trade-off with the execution time. To reach the best solution possible, the CP model and constructive heuristics were combined into a warm started CP model called \cpwseic\ which provides high-quality solutions even for short time limits and large instances. As demonstrated, \cpwseic\ can effectively solve instances of tens of machines and hundreds of tasks.

The efficiency of the proposed approaches was evaluated and compared against the existing state-of-the-art ones. Since, to the best of our knowledge, there is no paper tackling our considered $P|seq, ser|C_{max}$ problem, we compared our approaches to approaches for problems $PD|seq, ser=1|C_{max}$, $P|ser=1|C_{max}$ and $P|seq, ser=1|C_{max}$. The comparison was achieved by transforming the instances from original problems to our problem, however, this meant that our approaches had to tackle more complex instances. Despite that, our CP model proved to be better than all compared MIP models, providing better or equal solutions to every single instance tested and also being much better at proving the optimality. As for the comparison of \cpwseic\ to the heuristics for the compared problems, \cpwseic\ provided on par results against genetic algorithm for the most restricted problem $PD|seq, ser=1|C_{max}$ and better results than heuristics for $P|ser=1|C_{max}$ and for $P|seq, ser=1|C_{max}$. It also has the additional advantage of being able to reach optimum if enough time is provided.

In conclusion, we showed that our approaches are effective, relevant, and can be used for more specific problems with better efficiency than the existing approaches. We consider this the most significant contribution of this paper and see a great promise in applying proposed approaches in real-world production scenarios.

To accommodate even more real-world productions, we see great potential in introducing setups over setups to simulate operations like the movement of \cicmunda\ from one machine to another, as we did not see this addressed anywhere in the literature.

\section*{Acknowledgements}
This work was funded by EU Structural funds and the Ministry of Education, Youth and Sport of the Czech Republic within the project Cluster 4.0 number CZ.02.1.01/0.0/0.0/16\_026/0008432.
This work was supported by the EU and the Ministry of Industry and Trade of the Czech Republic under the Project OP PIK CZ.01.1.02/0.0/0.0/20\_321/0024399.

\newpage

\appendix

\section{Constructive Heuristic Pseudocodes}
\label{app:pseudocodes}

\subsection{Locally Optimal Selection of Setups (LOSOS)}

\singlespacing
\noindent \begin{minipage}{1.0\textwidth}
    \begin{algorithm}[H]
    \small
    \caption{Pseudocode of \lss\ algorithm.}
    \label{alg_losos}
    \begin{algorithmic}[1]
    \Function{\lss\ Solve}{}
        \State $T^{StartingTasks} \gets $ \Call{GenerateStartingTasks}{}\Comment{called function}
        \ForEach{$M_m \in M$}
            \State Schedule $M_m \gets T^{StartingTasks}_m$
        \EndFor
        \State $T^{Remaining} \gets T \setminus \{T^{StartingTasks}\}$
        
        \State $WorkersFreeFromTime \gets \{0\} \rightarrow |\workerSet|$ \Comment{List of zeroes of size $|\workerSet|$.}
        
        \While{$T^{Remaining} \neq \emptyset$}
            \State $ClosestEndingMachine \gets argmin(End(M))$ 
            \State $ClosestEndingTask \gets T_{-1, ClosestEndingMachine}$ \Comment{Last task is indexed by "-1".}
            
            \State $NextTask, \ SetupLength \gets $  \Call{SelectNextTask($ClosestEndingTask$, $T^{Remaining}$)}{}
            \State \Comment{called function}
            
            \State $Start(NextSetup) \gets max(End(ClosestEndingTask), min(WorkersFreeFromTime))$
            \State $Start(NextTask) \gets Start(NextSetup) + SetupLength$
            \State Schedule $ClosestEndingMachine \gets$ $NextSetup$
            \State Schedule $ClosestEndingMachine \gets$ $NextTask$ 
            
            \State argmin$(WorkersFreeFromTime) \gets Start(NextTask)$
            \State $T^{Remaining} \gets $ $T^{Remaining} \setminus \{ClosestEndingTask\}$
        \EndWhile
        \State \Call{OptimizeScheduleEnds}{}\Comment{called function}
    \EndFunction
    \end{algorithmic}
    \end{algorithm}
\end{minipage}

\doublespacing
Starting tasks $T^{StartingTasks}$ are selected, and one is assigned to every machine. There are multiple possible criteria for choosing $T^{StartingTasks}$; the best one is discussed in detail in subsection \ref{sec:constructive_heur:improv:start}. After these tasks are assigned, they are removed from the set of remaining tasks, which is denoted by $T^{Remaining}$. See lines 2 to 6.

Data structure representing all \cicmundove' availability is created, denoted by $WorkersFreeFromTime$. In algorithm implementation, the priority queue with the earliest available \cicmunda\ on top was used. At the start, all \cicmundove\ are considered ready to work; thus, their availability is set to time 0. See line 8.

The earliest ending task, denoted by $ClosestEndingTask$, and its respective machine, denoted by $ClosestEndingMachine$, are found. In algorithm implementation, a priority queue was used for storing the machines. Then, a suitable task, denoted by $NextTask$, and its respective setup length from $ClosestEndingTask$ are found. The methods of $NextTask$ selection are discussed in detail in \cref{sec:constructive_heur:improv}. See lines 11 to 14.

The earliest available \cicmunda\ is selected to perform the resulting $NextSetup$ between $ClosestEndingTask$ and $NextTask$. The setup's starting time is calculated according to \cicmunda's and $ClosestEndingMachine$'s availability and then is scheduled. After that, $NextTask$ is also scheduled to $ClosestEndingMachine$ and is removed from $T^{Remaining}$. See lines 16 to 22.

Finally, the end of the schedule is optimized by calling the \textsc{OptimizeScheduleEnds}. This is especially important if the default greedy $NextTask$ selection strategy was used, as it tends to leave long setups to the end. See line 25.


\subsection{Resolution of Setup Overlaps Lazily (ROSOL)}
The selection and assignment of $T^{StartingTasks}$ and the selection of $ClosestEndingTask$, $ClosestEndingMachine$ and $NextTask$ is the same as in \lss. The only difference here is that there is no data structure for \cicmundove. Thus, the $NextSetup$ and $NextTask$ are assigned to machines without considering \cicmunda\ availability. See lines 2 to 17.

The problem is now solved without considering \cicmundove' constraints. Before checking if it adheres to said constraints, \textsc{OptimizeScheduleEnds} is called to reduce makespan if possible. Calling \textsc{OptimizeScheduleEnds} is beneficial, even if the solution does not adhere to \cicmunda\ constraints, because machines' balancing will reduce future collisions when \textsc{OptimizeScheduleEnds} is called again at the end of the algorithm. See line 18.

All machines that are not in $SetMachines$ and currently have a setup scheduled are found and added to set $FreeMachines$. If needed, these machines' current setups can be moved to a later point in the schedule.  On the other hand, $SetMachines$ contains machines where the currently executed setup will not move under any circumstances as it has been already confirmed for the execution. $TimeStep$ is updated either by the end of the present task or setup on the machine. Also, if some fixed machine previously executing a setup is finished at current $Time$, it is removed from $SetMachines$. See lines 24 to 36.

If the current number of setups executed is higher than the number of \cicmundove, some setups from $FreeMachines$ will have to be moved to a later point in the schedule. The coefficient of move priority is computed based on the machine length compared to the makespan and the length of the currently present setup on the machine. This helps us determine how would the move affect the makespan and waiting times of other machines for free \cicmunda. See lines 37 to 42.

While free \cicmundove\ are available, the machine with the lowest coefficient is selected and moved to $SetMachines$, fixing its setup execution, thus allocating a \cicmunda\ to it. See lines 43 to 47.

When all \cicmundove\ are busy, and there are still setups currently being executed, they are moved to a later point in the solution together with their following tasks and setups on the same machine. The makespan of the solution is updated, and $Time$ is moved by $TimeStep$. See lines 48 to 53.

After resolution of the \cicmundove' collisions, the end of the schedule is again optimized. See line 57.


\singlespacing
\begin{algorithm}[H]
\small
\caption{Pseudocode of \rsl\ algorithm.}
\label{alg_rosol}
\begin{algorithmic}[1]
\Function{\rsl\ Solve}{}
    \State $T^{StartingTasks} \gets $ \Call{GenerateStartingTasks}{}\Comment{called function}
    \ForEach{$M_m \in M$}
        \State Schedule $M_m \gets T^{StartingTasks}_m$
    \EndFor
    \State $T^{Remaining} \gets $ $T \setminus \{T^{StartingTasks}\}$
    
    \While{$T^{Remaining} \neq \emptyset$}
        \State $ClosestEndingMachine \gets argmin(End(M))$ 
        \State $ClosestEndingTask \gets T_{-1, ClosestEndingMachine}$ \Comment{Last task is indexed by "-1".}
        \State $NextTask, \ SetupLength \gets $  \Call{SelectNextTask($ClosestEndingTask$, $T^{Remaining}$)}{}
        \State \Comment{called function}
        
        \State $Start(NextSetup) \gets End(ClosestEndingTask)$
        \State $Start(NextTask) \gets Start(NextSetup) + SetupLength$
        \State Schedule $ClosestEndingMachine \gets$ $NextSetup$
        \State Schedule $ClosestEndingMachine \gets$ $NextTask$ 
        
        \State $T^{Remaining} \gets $ $T^{Remaining} \setminus \{ClosestEndingTask\}$
    \EndWhile
    \State \Call{OptimizeScheduleEnds}{}\Comment{called function}
    
    \State $SetMachines \gets \emptyset$ \Comment{Machines fixed to execute current setup.}
    \State $Time \gets 0$ \Comment{$Time$ represents current time point in solution.}
    \While{$Time < Makespan$} \Comment{$Makespan$ represents solution's makespan.}
        \State $TimeStep \gets \infty$ \Comment{The amount of time to move.}
        \State $FreeMachines \gets \emptyset$ \Comment{Machines with movable current setup.}
        
        \ForEach{$M_m \in M$}
            \If{$M_m$ in $Time$ executes a setup}
                \State $FreeMachines \gets FreeMachines \bigcup \{M_m\}$
                \State $TimeStep \gets min(TimeStep, End(CurrentSetup_m) - Time)$
                \State \Comment{Currently executed setup given the current Time.}
            \Else
                \If{$M_m \in SetMachines$}
                    \State $SetMachines \gets SetMachines \setminus \{M_m\}$
                \EndIf
                \State $TimeStep \gets min(TimeStep, End(CurrentTask_m) - Time)$
                \State \Comment{Currently executed task given the current Time.}
            \EndIf
        \EndFor
        
        \If{$|FreeMachines| + |SetMachines| > |\workerSet|$}
            \ForEach{$M_m \in FreeMachines$}
                \State $SetupLength \gets End(CurrentSetup_m)- Start(CurrentSetup_m)$
                \State $MachineReserve \gets Makespan - End(M_m)$
                \State $Coefficient(M_m) \gets MachineReserve + SetupLength$
            \EndFor
        
            \While{$|SetMachines| < |\workerSet| \And |FreeMachines| > 0$} 
                \State $M_k \gets argmin(Coefficient(FreeMachines))$
                \State $FreeMachines \gets FreeMachines \setminus \{M_k\}$
                \State $SetMachines \gets SetMachines \bigcup \{M_k\}$
            \EndWhile
            
            \ForEach{$M_m \in FreeMachines$}
                \State Move tasks/setups starts/ends after current Time on $M_m$ by $TimeStep$
            \EndFor
            \State $Makespan \gets max(End(M))$
        \EndIf
        \State $Time \gets Time + TimeStep$
    \EndWhile

    \State \Call{OptimizeScheduleEnds}{}\Comment{called function}
\EndFunction
\end{algorithmic}
\end{algorithm}

\section{Large instance runtimes}
\label{app:runtimes}
\begin{figure}[!h]
    \centering
    \begin{tikzpicture}
    \begin{semilogyaxis}[
        width=1.0\textwidth,
        height=0.4\textwidth,
        legend cell align={left},
        legend pos=north west,
        legend columns=2,
        bar width=3.3pt,
        ybar=0pt,
        grid=both,
        scaled ticks = false,
        tick label style={
            /pgf/number format/fixed,
            /pgf/number format/precision=2
        },
        label style={font=\footnotesize},
        tick label style={font=\scriptsize},
        xtick=data,
        xticklabel style={rotate=90},
        enlargelimits=0.05,
        symbolic x coords={300x12000x30,300x12000x60,300x18000x30,300x18000x60,300x24000x30,300x24000x60,500x20000x50,500x20000x100,500x30000x50,500x30000x100,500x40000x50,500x40000x100}, 
        xticklabels={$300 \textbf{\text{\textbar}} 12000 \textbf{\text{\textbar}} 30$,$300\textbf{\text{\textbar}}12000\textbf{\text{\textbar}}60$,$300\textbf{\text{\textbar}}18000\textbf{\text{\textbar}}30$,$300\textbf{\text{\textbar}}18000\textbf{\text{\textbar}}60$,$300\textbf{\text{\textbar}}24000\textbf{\text{\textbar}}30$,$300\textbf{\text{\textbar}}24000\textbf{\text{\textbar}}60$,$500\textbf{\text{\textbar}}20000\textbf{\text{\textbar}}50$,$500\textbf{\text{\textbar}}20000\textbf{\text{\textbar}}100$,$500\textbf{\text{\textbar}}30000\textbf{\text{\textbar}}50$,$500\textbf{\text{\textbar}}30000\textbf{\text{\textbar}}100$,$500\textbf{\text{\textbar}}40000\textbf{\text{\textbar}}50$,$500\textbf{\text{\textbar}}40000\textbf{\text{\textbar}}100$}, 
        ylabel={Computation time $[ms]$},
        xlabel={}
    ]
    
        \addplot+[lightest] plot table [x=nmw, y=lss_avg, col sep=semicolon] {fig/app_huge_instances/execution_times_huge_instances.csv};
        \addplot+[mid] plot table [x=nmw, y=lss_se_avg2, col sep=semicolon] {fig/app_huge_instances/execution_times_huge_instances.csv};
        \addplot+[darkest] plot table [x=nmw, y=rsl_avg3, col sep=semicolon] {fig/app_huge_instances/execution_times_huge_instances.csv};
        \legend{\fontfamily{bch}\selectfont{\small{LOSOS}}, \fontfamily{bch}\selectfont{\small{LOSOS\smaller{$_{\text{SE}}$}}}, \fontfamily{bch}\selectfont{\small{ROSOL}}} 
    \end{semilogyaxis}
\end{tikzpicture}
    \caption{Runtimes of heuristics without expensive (task coefficient and idleness reduction) improvements. The x-axis labels are in format \machineLastIndex\ \textbf{\text{\textbar}} $t$ \textbf{\text{\textbar}} $\workerLastIndex$.}
    \label{fig:runtimes}
\end{figure}
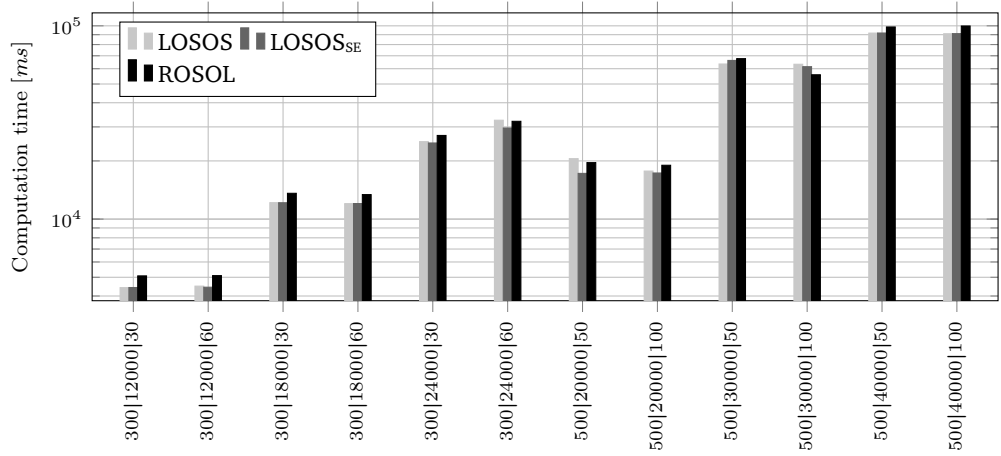

\section{Lower bound calculation}
\label{app:lower_bound}
The lower bound calculation to assess overall quality of our proposed approaches is following:
\begin{itemize}
    \item First, $\overline{M}_{p} = \sum_{i=1}^{t} \frac{p_i}{m}$ is calculated.
    \item For every task \taskTwo, $z_j = \min_{T_i \in T}(\setupTimeEq)$, i.e., the shortest setup from any \taskOne $\in$ \taskSet\ to a given \taskTwo, are calculated and added to multiset $Z$. Next, $m$ largest $z_j$ are removed from $Z$, yielding $Z^\prime$. 
    Then, $\overline{M}_{o} = \sum_{z \in Z^\prime} \frac{z}{m}$ and $\overline{R}_{o} = \sum_{z \in Z^\prime} \frac{z}{r}$ are calculated.
    \item Finally, $lower bound = max(\overline{M}_{p} + \overline{M}_{o}, \overline{R}_{o})$. Please note that $\overline{M}_{p} + \overline{M}_{o}$ represent smallest possible time that the longest machine will take to execute the schedule, while $\overline{R}_{o}$ smallest possible time that the longest working \cicmunda\ will take to perform the setups in the schedule.
\end{itemize}

\section{CP scalability}
\label{app:cp_scalability}
\begin{table}[H]
    \centering
    \scalebox{0.85}{
    \begin{tabular}{lllllll}
        \hline
        $\#$ & $m$ & $t$ & $r$ & \lsseic & \rsle & \cpwseic (3600s) \\
        \hline
         1             & 30           & 600           & 3            & 568                   & 548                 & \textbf{521}         \\
         2             & 30           & 600           & 6            & 546                   & 548                 & \textbf{522}         \\
         3             & 30           & 750           & 3            & 676                   & 672                 & \textbf{647}         \\
         4             & 30           & 750           & 6            & 668                   & 672                 & \textbf{647}         \\
         5             & 40           & 800           & 4            & 555                   & 557                 & \textbf{534}         \\
         6             & 40           & 800           & 8            & 560                   & 554                 & \textbf{533}         \\
         7             & 40           & 1000          & 4            & 703                   & 689                 & \textbf{668}         \\
         8             & 40           & 1000          & 8            & 689                   & 689                 & \textbf{668}         \\
         9             & 50           & 1000          & 5            & 565                   & 559                 & \textbf{537}         \\
         10            & 50           & 1000          & 10           & 556                   & 559                 & \textbf{535}         \\
         11            & 50           & 1250          & 5            & 707                   & 710                 & \textbf{678}         \\
         12            & 50           & 1250          & 10           & 697                   & 710                 & \textbf{681}         \\
         \hline
         $\sum$        & -            & -             & -            & 7490                  & 7467                & \textbf{7171}        \\
         $RD$          & -            & -             & -            & 4.45 \%                & 4.13 \%              & \textbf{0.0 \%}       \\
        \hline
    \end{tabular}
}
    \caption{The comparison between the warm starts (\lsseic\ and \rsle) provided to \cpwseic\ and solutions attained by \cpwseic\ in an hour runtime.}
    \label{tab:cp_scalability}
\end{table}

\bibliographystyle{elsarticle-num} 
\bibliography{cas-refs}






\end{document}